\documentclass{article}

\usepackage{arxiv}

\usepackage[utf8]{inputenc} % allow utf-8 input
\usepackage[T1]{fontenc}    % use 8-bit T1 fonts
\usepackage{hyperref}       % hyperlinks
\usepackage{url}            % simple URL typesetting
\usepackage{booktabs}       % professional-quality tables
\usepackage{amsfonts}       % blackboard math symbols
\usepackage{nicefrac}       % compact symbols for 1/2, etc.
\usepackage{microtype}      % microtypography
\usepackage{lipsum}
\usepackage{graphicx}

\usepackage{amssymb}
\usepackage{amsmath}
\usepackage{amsthm}

\newtheorem{theorem}{Theorem}

\newtheorem{definition}{Definition}
\usepackage[]{rotating}

\usepackage[linesnumbered,lined,boxed,commentsnumbered]{algorithm2e}
\usepackage{caption}

\title{Component-wise Adaptive Trimming For Robust Mixture Regression}

\author{
 Wennan Chang\\
  Electrical and Computer Engineering\\
  Purdue University\\
  West Lafayette, IN \\
  \texttt{chang534@purdue.edu} \\
   \And
 Xinyu Zhou \\
  Department of Computer Science\\
  Indiana University\\
  Bloomington, IN \\
  \texttt{zhou19@iu.edu} \\
  \And
 Yong Zang \\
  Department of Biostatistics\\
  Indiana University School of Medicine\\
  Indianapolis, IN 46202 \\
  \texttt{zangy@iu.edu} \\
   \And
   Chi Zhang* \\
   Electrical and Computer Engineering \\
   Purdue University \\
   \texttt{czhang87@iu.edu} \\
   \And
   Sha Cao* \\
   Department of Biostatistics \\
   Indiana University School of Medicine \\
   \texttt{shacao@iu.edu} \\

}

\begin{document}
\maketitle
\begin{abstract}
Parameter estimation of mixture regression model using the expectation maximization (EM) algorithm is highly sensitive to outliers. Here we propose a fast and efficient robust mixture regression algorithm, called \textbf{C}omponent-wise \textbf{A}daptive \textbf{T}rimming (\textbf{CAT}) method. We consider simultaneous outlier detection and robust  parameter estimation to minimize the effect of outlier  contamination. Robust mixture regression has many important applications including in human cancer genomics data, where the population often displays strong heterogeneity added by unwanted technological perturbations. Existing robust mixture regression methods suffer from outliers as they either conduct parameter etimation in the presence of outliers, or rely on prior knowledge of the level of outlier contamination. CAT was implemented in the framework of classification expectation maximization, under which a natural definition of outliers could be derived. It implements a least trimmed squares (LTS) approach within each exclusive mixing component, where the robustness issue could be transformed from the mixture case to simple linear regression case. The high breakdown point of the LTS approach allows us to avoid the pre-specification of trimming parameter. Compared with multiple existing algorithms, CAT is the most competitive one that can handle and adaptively trim off outliers as well as heavy tailed noise, in different scenarios of simulated data and real genomic data. CAT has been implemented in an R package `RobMixReg' available in CRAN.
\end{abstract}

% keywords can be removed
% \keywords{First keyword \and Second keyword \and More}

\section{Introduction}
Finite Mixture Gaussian Regression (FMGR) model was first introduced by Goldfeld et al \cite{goldfeld1973estimation}, and has been widely used to explore the latent relationship between a response and independent variables in many fields \cite{bohning1999computer,hennig2000identifiablity,jiang1999hierarchical,mclachlan2004finite,xu1996convergence,fruhwirth2006finite}. Parameter estimation in FMGR is usually conducted through maximum likelihood using expectation maximization (EM) algorithm assuming normally distributed component errors, which is vulnerable to outliers or heavy-tailed noises. Many algorithms have been developed to estimate the FMGR parameters robustly \cite{yu2020selective}. Using the idea of weighted regression, Markatou \cite{markatou2000mixture} and Shen et al. \cite{shen2004outlier} proposed using a weight factor for each data point to robustify the estimation procedure. By modifying the M-step in the EM algorithm, Bai et al. \cite{bai2012robust} replaced the least squares criterion in M step with a robust bi-square criterion (\textbf{MIXBI}); Bashir and Carter \cite{bashir2012robust} adopted the idea of the S-estimator to mixture of linear regression;  Song et al. \cite{song2014robust} proposed using Laplace distribution to model the error distribution (\textbf{MIXL}); Yao et al. \cite{yao2014robust} extended the idea of mixture of \textit{t}-distributions proposed by Peel and McLachlan \cite{peel2000robust} from clustering to the regression setting (\textbf{MIXT}). These methods seek for robust parameter estimation in the presence of outliers, however, the outliers may still corrupt the robust algorithms, and the identities of the outliers still remain unknown unless further screening steps are taken. The identities of the outliers are often interesting for two reasons: firstly, removal of the outliers could reduce their effect on the estimators and improve the estimation accuracy; secondly, for practical reasons, outlying samples could be caused by measurement errors, or they may represent a novel mechanism not representative by the majority of the current observations, both of which are worthy of further investigation. 

Currently, to enable outlier detection, usually a hyperparameter regarding the proportion of outlying samples needs to be specified. Yu et al. \cite{yu2017new} proposed a penalized mean-shift mixture model, $RM^2$, for simultaneous outlier detection and robust parameter estimation.
Neykov et al. \cite{neykov2007robust} proposed the trimmed likelihood estimator (\textbf{TLE}), where given a trimming parameter $\alpha$, $0 \leq \alpha \leq 1$, the outliers are defined as the $N \alpha$ observations with the smallest sample likelihood, and they presented a Trimmed Likelihood Estimator (TLE) algorithm based on EM algorithm, and a FAST-TLE algorithm using classification EM algorithm \cite{celeux1992classification}. Similar to fast-TLE, Dogru and Arslan \cite{dougru2018robust} proposed the adapted complete data log-likelihood function using the least trimmed squares (LTS)  \cite{rousseeuw1984least} criterion, where the sum of log-likelihood for portions of the data points were optimized. Following very similar steps as TLE algorithm, García-Escudero et al. \cite{garcia2010robust, garcia2017robust} proposed an algorithm (\textbf{CWM}) with further control over the scattering parameters, as well as a second trimming strategy on the explanatory variables. The challenge with the trimming based algorithms are the involvement of hyperparameters, namely,  penalty parameter in $RM^2$ and the trimmning parameter $\alpha$ in the other algorithms, which could heavily impact the performance of these trimming based algorithms. Yu et al. \cite{yu2017new} proposed using BIC procedure for hyperparameter tuning, however, BIC criterion becomes highly unstable when the total number of parameters, which equals to the total number of outliers, becomes large. As discussed in a recent review article \cite{yu2020selective}, the involved parameters are interrelated with the number of components, where a high trimming level will lead to the removal of components with fewer observations. In summary, there is a lack of an algorithm that could adaptively trim the outlying samples to minimize its impact on the parameter estimation, while avoiding the pre-specification of the level of trimming.

In biomedical science, the patient population often consists of different molecular subtypes \cite{guinney2015consensus}, and the outlying samples introduced by technological errors makes it more challenging to tease out the latent relations among the genomic markers using FMGR. To address the challenges in simultaneous outlier detection and robust parameter estimation in FMGR, we adopted the idea of Classification-Expectation-Maximization (CEM) algorithm, where individual observations are assigned to a definite cluster as part of the maximization process, different from traditional EM algorithm \cite{celeux1992classification}. Essentially, CEM maximizes the complete data likelihood, instead of the observed data likelihood as in traditional EM, and has been shown to outperform traditional EM with faster convergence rate and better or comparable estimates  \cite{faria2010fitting}. Under CEM, each component has its exclusive members, which makes it possible to apply a trimmed likelihood approach designed for (single component) linear regression on its member, and hence enables both robust parameter estimation as well as outlier detection for the component. Our major contribution in this paper is, by introducing CEM to FMGR, we provided a platform that migrates the robustness issue from mixture regression to (single component) linear regression, for which LTS estimators have been extensively studied \cite{siegel1982robust,rousseeuw1984least}. Therefore, since LTS-based robust regression has a high breakdown point, we could avoid the pre-specification of trimming parameter, by simply opt to maximize the sum of decreasing ranked likelihood of a sufficiently small portion of the samples within each component, namely, 0.5. In addition, since the task of outlier detection in the mixture model was converted to that of linear regression in each component, it is possible to formally define outliers in FMGR. Overall, our algorithm detects outlier in a data-driven fashion free of hyperparameters, and is hence computationally efficient and user-friendly.

The remainder of the article is organized as follows. In Section 2, we will introduce the complete data maximum likelihood, and the CEM algorithm, based on which, our component-wise adaptive trimming method is developed. In Section 3, we show the performance comparison of our method with other six state of the art methods on synthetic datasets. In section 4, we will apply all methods to a real world dataset studying the heterogeneous DNA methylation regulatory effects on gene expression in colon cancer.

\section{Component-wise adaptive trimming}
\subsection{The complete data maximum likelihood estimation}

Let $Y=(y_1,...,y_N)^T \in \mathcal{R}^N$, $X= [ \boldsymbol{x}_1,...,\boldsymbol{x}_N ]^T \in \mathcal{R}^{N \times (P+1)}$ be a finite set of observations, and $X$ the design matrix, and $Y$ the response vector. Consider an FMGR model parameterized by $\boldsymbol\theta = \{ (\pi_k, \boldsymbol{\beta}_k, \sigma_k^2) \}_{k=1}^K$, it  is assumed that when $(\boldsymbol{x}, y)$ belongs to the \textit{k}-th component, $k=1,...,K$, then $y = \boldsymbol{x}^T \boldsymbol{\beta}_k + \epsilon$, where $\epsilon \sim N(0, \sigma_k^2)$. Then, the conditional density of $y$ given $\boldsymbol{x}$ is $f(y| \boldsymbol{x,\theta}) = \sum_{k=1}^K \pi_k \mathcal{N}(y; \boldsymbol{x^T \beta_k}, \sigma_k^2)$, where $\mathcal{N}(y; \mu, \sigma^2 )$ is the normal density function with mean $\mu$ and variance $\sigma^2$. Let $z$ be the membership indicator for observation $(\boldsymbol{x}, y)$, then $\pi_k = p(z=k)$. The maximum likelihood estimate for $\boldsymbol\theta$ is through maximizing the following log likelihood:
% \begin{linenomath*}
\begin{equation}
\mathcal{L}_{X,Y} (\boldsymbol{\theta}) := \sum_{i=1}^N \log(\sum_{k=1}^K \pi_k \mathcal{N} (y_i; \boldsymbol{x_i^T \beta_k}, \sigma_k^2))
\end{equation}
% \end{linenomath*}
EM algorithm is usually applied to obtain the MLE estimates, by treating the cluster membership $z$ as missing random variables. %However, such an algorithm is not natural for defining outliers, as we will discuss later. 

Assume, we are given a set of observations $(\boldsymbol {x}_i,y_i )_{i=1}^N$ and assignments $(z_i )_{i=1}^N$. Then, the likelihood that all observations have been drawn according to a FMGR $\boldsymbol{\theta}$ and that each observation $(\boldsymbol{x}_i,y_i)$ has been generated by the $z_i$-th component, is given by 
% \begin{linenomath*}
\begin{equation}
\prod_{i=1}^N p(y_i, z_i | \boldsymbol{x}_i, \boldsymbol{\theta}) = \prod_{i=1}^N \pi_{z_i} \mathcal{N} (y_i; \boldsymbol{x_i^T \beta}_{z_i}, \sigma_{z_i}^2)
\end{equation}
% \end{linenomath*}

This is called the complete-data likelihood. Note that the assignments $\{z_i\}_{i=1}^N$ define a partition of the $N$ observations, $\mathcal{C} = \bigcup_{k=1}^K C_k$, such that $i \in C_k$ iff $z_i = k$. Denote $n_k$ as the total number of elements in $C_k$, we can then rewrite the Equation (2) in its logarithm form as
% \begin{linenomath*}
\begin{equation}
\mathcal{L}_{X,Y}^f (\boldsymbol{\theta, \mathcal{C}}) := \sum_{k=1}^K \{\sum_{i\in C_k} \log\mathcal{N} (y_i;\boldsymbol{x}_i^T \boldsymbol{\beta}_k,\sigma_{k}^2 ) + \log \pi_k n_k\}
\end{equation}
% \end{linenomath*}
We introduce the complete data maximum likelihood estimates (CMLE) as follows.

\begin{definition}{\textbf{(Complete-data Maximum Likelihood Estimates, CMLE)}}
Let $X$ be the design matrix, and $Y$ be the response vector. Given an integer $K$, find a partition $\mathcal{C} = \{ C_1,...,C_K \}$ of the N observations and FMGR parameters $\boldsymbol{\theta} = \{ (\pi_k, \boldsymbol{\beta}_k, \sigma_k) \}_{k=1}^K$ that maximizes $\mathcal{L}^f_{X,Y} (\boldsymbol{\theta}, \mathcal{C})$ defined in equation (3).
\end{definition}

Note that, CMLE is not well defined in this form. For example, for an observation $(\boldsymbol{x}_i, y_i)$, if $\boldsymbol{\beta_k}$ is chosen such that $y_i = \boldsymbol{x}_i^T \boldsymbol{\beta}_k$ and we let $\sigma_k \rightarrow{0}$, then $f(y_i; \boldsymbol{x}_i, \boldsymbol{\theta}) \rightarrow{\infty}$, which results in infinite likelihood. A common practice is to put some mild restrictions on the cluster size or the variance parameter, then we can lower bound the variance associated with each regression line, and the CMLE will be well defined \cite{blomer2016hard, garcia2017robust}.

\subsection{Alternating Optimization Scheme with the CEM algorithm}

We introduce the alternating optimization algorithm to solve the CMLE problem \cite{blomer2016hard}. Clearly, fixing the partition $\mathcal{C} = \{ C_1,...,C_K \}$, the optimal mixture parameter is given by $\boldsymbol{\theta} = \{ (\pi_k, \boldsymbol{\beta}_k, \sigma_k) \}_{k=1}^K$ with
% \begin{linenomath*}
\begin{equation}
\pi_k = \frac{n_k}{\sum_{l=1}^K n_l} 
\end{equation}
% \end{linenomath*}
% \begin{linenomath*}
\begin{equation}
(\boldsymbol{\beta}_k, \sigma_k^2) = \textbf{OLS}(Y_{C_k}, X_{C_k,;})
\end{equation}
% \end{linenomath*}
Here, $\textbf{OLS}({Y}_{C_k}, X_{C_k,;})$ means the ordinary least squares solution to regressing $Y$ on $X$ using only observations from $C_k$.

Fixing the FMGR parameters $\boldsymbol{\theta} = \{ (\pi_k, \boldsymbol{\beta}_k, \sigma_k) \}_{k=1}^K$, the optimal partition is given by assigning each point to its most likely component, i.e.
% \begin{linenomath*}
\begin{equation}
i \in C_k \Longleftrightarrow k = \underset{l \in \{1,...,K \}}{\mathbf{argmax}} p(z_i=l | \boldsymbol{x}_i, y_i, \boldsymbol{\theta})
\end{equation}
% \end{linenomath*}
where
% \begin{linenomath*}
\begin{equation}
p(z_i=k | \boldsymbol{x}_i, y_i, \boldsymbol{\theta}) = \frac{\pi_k \mathcal{N}(y_i;\boldsymbol{x}_i^T \boldsymbol{\beta}_k, \sigma_k^2)}{\sum_{l=1}^K \pi_l \mathcal{N}(y_i; \boldsymbol{x}_i^T \boldsymbol{\beta}_l, \sigma_l^2)}
\end{equation}
% \end{linenomath*}
which is the posterior probability that $(\boldsymbol{x}_i, y_i)$ lies on the \textit{k-}th regression line of the mixtures. By repeatedly updating between $\boldsymbol{\theta}$ and $\mathcal{C}$, we will show in Theorem 1 that the solution converges to a stationary point of the full likelihood function. We call this alternating scheme the CEM algorithm, and Algorithm 1 outlined the major steps including initialization, estimation, classification and maximization steps.

\begin{algorithm}
\SetKwInput{kwInit}{Initialization}
\DontPrintSemicolon
\KwIn{Response vector ${Y}$; independent variables in matrix $X_{N\times(P+1)}$; the number of mixing component, $K$; size of initialization random sample, $n_0$; the maximum number of iteration $L_0$}
\KwOut{$\boldsymbol{\theta}=\{ \pi_k, \boldsymbol{\beta}_k \}_{k=1}^K; \mathcal{C} = \bigcup_{k=1}^K \mathcal{C}_k $}

% \SetKwBlock{Begin}{For}{end For}
% \Begin($k=1,...,K$)
\For{$k=1,...,K$}{
  Draw a random sample of size $n_0$ from set $\{1,...,N\}$, indexed by $I_k$ \;
  Run ordinary linear regression to get initial regression parameter estimates: $( \boldsymbol{\beta}_k^{(0)}, \sigma_k^{(0)})=: \textbf{OLS} ( {Y_{I_k} \sim X_{I_k,:}} ) $ \;
}

% \SetKwBlock{Begin}{For}{end For}
% \Begin($m = 0,..., L_0$ or until convergence)
\For{$m = 0,..., L_0$ or until convergence}{
  \textbf{E-step}: Compute for $i=1,...,N$ and $k=1,...,K$, the current posterior probabilities $p_{ik}^{(m)}$ by \; $p_{ik}^{(m)} = p(z_i=k | \boldsymbol{x_i}, y_i, \boldsymbol{\theta}^{(m)})$ \;
  
  \textbf{C-step}: For $k=1,...,K$, assign $C_k^{(m)}= \{ i| \underset{l \in \{ 1,...,K \}}{\operatorname{argmax}} \ p_{il}^{(m)}=k, i=1,...,N\}$, and let $n_k^{(m)}$ be the size of $C_k^{(m)}$  \;
  
  \textbf{M-step}: For $k=1,...,K$, the parameters are then updated by 
  $\pi_k^{(m+1)} = \frac{n_k^{(m)}}{\sum_{l=1}^K  n_l^{(m)} }$, \;
  $( \boldsymbol{\beta}_k^{(m+1)}, \sigma_k^{2(m+1)}) = 
  \textbf{OLS} (  {{Y}_{C_k^{(m)}}, X_{C_k^{(m)},:}} )$  
}

\caption{CEM}\label{CEM}
\end{algorithm}

% \begin{algorithm}
%   \caption{CEM}
%   \begin{algorithmic}
%     \State Input: $X_{N \times P}, Y_{N \times 1}, K$
%     \State Initialization: $\boldsymbol{\theta}^{(0)}$
%     \State For $m = 0,..., Max \ Iteration$
%     \State (1) E-step: Compute for $i=1,...,N$ and $k=1,...,K$, the current posterior probabilities $p_{ik}$ by
% \begin{equation*}
% p_{ik} = p(z_i=k | \boldsymbol{x_i}, y_i, \boldsymbol{\theta}^{(m)})
% \end{equation*}
%     \State (2) C-step: For $k=1,...,K$, assign $C_k^{(m+1)}= \{ i| \underset{l \in \{ 1,...,K \}}{\operatorname{argmax}} \ p_{il}=k, i=1,...,N\}$.
%     \State (3) M-step: For $k=1,...,K$, the parameters are then updated by
% \begin{equation*}
% \pi_k^{(m+1)} = \frac{|\mathcal{C}_k^{(m+1)}|}{\sum_{l=1}^K | C_l^{(m+1)} |}
% \end{equation*}
% \begin{equation*}
% ( \boldsymbol{\beta}_k^{(m+1)}, \sigma_k^{2(m+1)}) = 
% \Call{Ols}{{Y}_{C_k^{(m+1)}}, X_{C_k^{(m+1)},:}}
% \end{equation*}
%     \State \indent Stop if converged.
%     \State End

% %    \\\hrulefill
%     \State Output: $\boldsymbol{\theta}: \mathcal{C} = \bigcup_{k=1}^K \mathcal{C}_k, \{ \pi_k, \boldsymbol{\beta}_k \}_{k=1}^K$
%   \end{algorithmic}
% \end{algorithm}
% \algcomment{\textbf{Note}: Here $( {Y}_{C_K^{(m+1)}}, X_{C_k^{(m+1)},:} )$ denotes the observations indexed by $C_k^{(m+1)};$ .}

\begin{theorem}
The complete data log likelihood, $\mathcal{L}^f_{X,Y} (\boldsymbol{\theta}^{(m)}, \mathcal{C}^{(m)})$, is non-decreasing for any sequence $\mathcal{C}^{(m)}, \boldsymbol{\theta}^{(m)}$ defined as in Algorithm 1, and it converges to a stationary value. Moreover, if the maximum likelihood estimates of the parameters are well-defined, the sequence of $\mathcal{C}^{(m)}, \boldsymbol{\theta}^{(m)}$ converges to a stationary position. \\
\label{theorem:1}
\end{theorem}

The CEM algorithm has been popularly used in both the clustering and regression-based clustering settings. Obviously, it is vulnerable to outliers, and in the next sections, we introduce our robust procedure on top of the CEM algorithm.

\subsection{A new definition for outlier under CEM}
In linear regression, outliers are understood as observations that deviate from the model assumptions, and obviously, samples with lower likelihood are more likely to be outliers. If the ratio of outliers, $\alpha$, is known, the outliers are identified as the ratio $\alpha$ of the total observations with the lowest likelihood.

Unfortunately, such a definition for outliers becomes less applicable in the case of mixture regression. Given a robust mixture regression model and a trimming ratio $\alpha$, if we follow the same logic as in linear regression, then the $\lceil n * \alpha \rceil$ observations with the smallest overall likelihood will be detected as outliers, as in \cite{neykov2007robust}. This trimmed likelihood approach implies that an observation with lower overall likelihood is more likely to be an outlier than an observation with higher overall likelihood. However, the overall likelihood depends on not only the likelihood of the observation with respect to each component, but also the proportion of each component, and such a criteria for outlier becomes problematic if the mixing components are unbalanced. In other words, a low $\pi_k$ will down-weigh the ``outlierness" of an observation from the \textit{k-}th component. In addition, if we argue that, given a set of observations, we could always find certain mixture model to well explain it, there is no basis for us to call any observation an outlier.

The complete data likelihood approach based CEM algorithm disentangles the mixture distribution into exclusive clusters, within which, the robustness issue could be much easily handled give the tremendous amount of research conducted for robust linear regression. More importantly, we could introduce a more natural definition for outliers.

\begin{definition}{\textbf{(Outliers of FMGR)}}
Given an FMGR model parameterized by $\boldsymbol{\theta} = \{ (\pi_k, \boldsymbol{\beta}_k, \sigma_k) \}_{k=1}^K$, under CMLE, an observation $(\boldsymbol{x}_i, y_i)$ is considered as an outlier, if $i \in \mathcal{C}_k$ and $|y_i - \boldsymbol{x}_i^T\boldsymbol\beta_k| \geq \eta_k (\sigma_k)$. In other words, an observation is considered as an outlier if it is an outlier to the component it belongs to. 
\end{definition}

\noindent Here $\eta_k(\cdot)$ is a criteria for outlier-ness in linear regression, which usually depends on the variance level of the component. This new definition transforms the robustness issue from a mixture model to its $K$ linear regression components, the latter of which has been well defined and studied. Different from the overall likelihood-based outlier definition adopted by TLE and CWM, our definition of outlier does not involve the cluster prior, and is hence more fair to clusters with relatively smaller sizes. In fact, in supplementary Tables S1-S4, we demonstrated using simulation data that, under unbalanced cluster sizes, TLE and CWM, which defines outliers based on the overall likelihood, perform much worse than our proposed method, which adopted our new definition of outliers. 

Naturally, to confer a robust parameter estimation for FMGR under CMLE, we could replace the least square criterion for parameter estimation in the M-step by a robust criterion; and further to enable simultaneous outlier detection, we could choose to use any trimmed likelihood approach with high break-down point \cite{donoho1983notion}.  

\subsection{The robust CEM algorithm}
\noindent Under Definition 2, detecting outliers of the FMGR model could be accomplished through detecting the component-wise outliers. Many robust estimators have been proposed for linear regression to achieve high breakdown point or high efficiency or both \cite{yu2017robust}, where the objective of minimizing the sum of the squared residuals has been replaced by more robust measures. Among them, the Least Median of Squares (LMS) estimates \cite{siegel1982robust,rousseeuw1984least} which minimize the median of squared residuals, Least Trimmed Squares (LTS) estimates \cite{rousseeuw1984least,rousseeuw2005robust} which minimize the trimmed sum of squared residuals, and S-estimates \cite{rousseeuw1984robust} which minimize the variance of the residuals, remain to be powerful robust algorithms with a breakdown point as high as 0.5, the best that can be expected. This means that the resulting estimators from these algorithms can resist the effect of nearly 50\% of contamination in the data. 

To achieve simultaneous outlier detection and robust parameter estimation, instead of maximizing the component-wise sum of likelihood in CEM, our \textbf{C}omponent-wise \textbf{A}daptive \textbf{T}rimming method, namely \textbf{CAT}, maximizes the component-wise sum of trimmed likelihood, i.e.,

% \begin{linenomath*}
\begin{equation}
\mathcal{L}_{X,Y}^{f,trim} (\boldsymbol{\theta, \mathcal{C}}) := \sum_{k=1}^K \sum_{i=g_k}^{h_k}\{ (l_{k})_{i:n} + \log \pi_k \}
\end{equation}
% \end{linenomath*}

where $l_{ki}=\log\{\mathcal{N} (y_i;\boldsymbol{x}_i^T \boldsymbol{\beta}_k ,\sigma_{k}^2)\}$, and $(l_{k})_{r:n_k}$ denotes the $r$-th largest value of the sequence of $l_{ki}$, namely, $(l_{k})_{1:n_k}\geq \cdots \geq (l_{k})_{n_k:n_k} $. For component-wise LMS estimates, one could let $g_k=h_k=[n_k/2]+1$; for LTS estimate, one could let $g_k=1, h_k=[n_k/2]+1$, which obtains the highest break-down point of LTS. Our CAT algorithm adopted the latter robust procedure in order to avoid the selection of a trimming parameter. The high breakdown point of the LTS algorithm makes it possible to minimize the effect of outliers in parameter estimation even if only half of the likelihood were optimized, and we could thus develop a data-driven algorithm for simultaneous outlier detection and robust parameter estimation in FMGR. 

In general, as outlined in Algorithm 2, our \textbf{CAT} algorithm implements very similar steps to the CEM algorithm outlined in Algorithm 1, except that the OLS estimates in the M-step is replaced by the LTS estimate. CAT starts by initializing the posterior probability matrix, $W$. For $k=1,...,K$, we randomly draw $n_0$ samples to build a robust linear regression model, and the posterior probability of sample $i$ for component $k$ will be initialized as the density of sample $i$ fitting the $k$-th robust regression line. For robust linear regression with trimmed likelihood approach, we used the ``ltsReg" function in the ``robustbase" library in R \cite{pison2002small, leroy1987robust, rousseeuw1984least, rousseeuw1999fast}, where the parameters were estimated to maximize the sum of the likelihood of the largest half, and outliers were detected as those with relatively large residuals. With initialized $W$, CAT then runs a robust CEM algorithm where the OLS estimates in Algorithm 1 in the M-step was replaced by robust estimates using trimmed likelihood method. 

\begin{algorithm*}
\SetKwInput{kwInit}{Initialization}
\DontPrintSemicolon
\KwIn{Response vector ${Y}$; independent variables in matrix $X_{N\times(P+1)}$; the number of mixing component, $K$; size of initialization random sample, $n_0$; the maximum number of iteration $L_0$}
\KwOut{Robust FMGR parameter estimate $\boldsymbol\theta^* = \boldsymbol\theta$; outlier set $U^*=U$}
\kwInit{Same as Algorithm 1}

% \SetKwBlock{Begin}{For}{end For}
% \Begin($m = 0,..., L_0$ or until convergence)
\For{$m = 0,..., L_0$ or until convergence}{
     
  \textbf{E-step}: Compute for $i=1,...,N$ and $k=1,...,K$, the current posterior probabilities $p_{ik}^{(m)}$ by \; $p_{ik}^{(m)} = p(z_i=k | \boldsymbol{x_i}, y_i, \boldsymbol{\theta}^{(m)})$ \;
  
  \textbf{C-step}: For $k=1,...,K$, assign $C_k^{(m)}= \{ i| \underset{l \in \{ 1,...,K \}}{\operatorname{argmax}} \ p_{il}^{(m)}=k, i=1,...,N\}$, and let $n_k^{(m)}$ be the size of $C_k^{(m)}$  \;
  
  \textbf{M-step}: For $k=1,...,K$, run robust linear regression using samples in $C_k^{(m)}$, i.e.,   $( \boldsymbol{\beta}_k^{(m+1)}, \sigma_k^{2(m+1)}) = 
  \textbf{RLM} (  {{Y}_{C_k^{(m)}}, X_{C_k^{(m)},:}} )$ ;   $\pi_k^{(m+1)} = \frac{n_k^{(m)}}{\sum_{l=1}^K  n_l^{(m)} }$;and let $U_k$ be outliers of component $k$
     }

\caption{CAT: Component-wise Adaptive Trimming}\label{CAT: Component-wise Adaptive Trimming}
\end{algorithm*}

\begin{algorithm*}
\DontPrintSemicolon
\SetKwInput{kwInit}{Initialization}
\KwIn{Response vector ${Y}$; independent variables in matrix $X_{N\times(P+1)}$; the number of mixing component, $K$; size of initialization random sample, $n_0$; the maximum number of iteration $L_0$}
\KwOut{Robust FMGR parameter estimate $\boldsymbol\theta^* = \boldsymbol\theta$; outlier set $U^*=U$}
\kwInit{Same as Algorithm 1}
% \SetKwBlock{Begin}{For}{end For}
% \Begin($m = 0,..., L_0$ or until convergence)
\For{$m = 0,..., L_0$ or until convergence)}{

\textbf{E-step}: Same as Algorithm 2 \;
\textbf{C-step}: Same as Algorithm 2 \;
  
\textbf{M-step}: For $k=1,...,K$, run robust linear regression using samples in $C_k^{(m)}$, i.e., $\textbf{RLM} (  {{Y}_{C_k^{(m)}}, X_{C_k^{(m)},:}} )$; and let $U_k$ be outliers of component $k$ \;

\textbf{Refit-step}:
Let $U=\bigcup_k U_k;S=\{1,...,N\}-U$; let $\boldsymbol{\theta}_{S}$ be the MLE estimates of fitting $K$ mixture regression lines using samples in $S$ only, and update the parameter $\boldsymbol{\theta}^{(m+1)})=\boldsymbol{\theta}_{S}$ \;

}
\caption{fast-CAT: Component-wise Adaptive Trimming}\label{fast-CAT: Component-wise Adaptive Trimming}
\end{algorithm*}

\begin{theorem}
The component-wise trimmed complete data log likelihood, $\mathcal{L}^{f,trim}_{X,Y} (\boldsymbol{\theta}^{(m)}, \mathcal{C}^{(m)})$,
is non-decreasing for any sequence $\mathcal{C}^{(m)}, \boldsymbol{\theta}^{(m)}$ defined as in Algorithm 2, and it converges to a stationary value. Moreover, if the component-wise trimmed maximum likelihood estimates of the parameters are well-defined, the sequence of $\mathcal{C}^{(m)}, \boldsymbol{\theta}^{(m)}$ converges to a stationary position. \\
\label{theorem:2}
\end{theorem}

In the M-step, CAT updates the parameter $\boldsymbol{\theta}$ using the robust estimates of each component. We als propose a fast implementation, fast-CAT, to achieve faster convergence. As outlined in Algorithm 3, CAT adopted a model refit step, where an MLE estimate of $\boldsymbol{\theta}$ is obtained using the non-outlying samples only, as an update of the $\boldsymbol{\theta}$ at the current step, instead of the component-wise LTS estimates used in Algorithm 2. Note that standard regression analysis tools can be applied to recover the observations that should not have been regarded as outliers. The MLE estimates were conducted using function ``flexmix" from the ``flexmix" R package \cite{leisch2004flexmix}. Similar to other algorithms, in our package implementation, we used multiple random starts to stabilize the results.

\section{Simulation studies}

We evaluated the performance of CAT, using the fast-CAT implementation in Algorithm 3, on synthetic datasets, and compare it with several existing method, including MLE, TLE, CWM, MIXBI, MIXL, and MIXT. They stand for the maximum likelihood estimates using traditional EM algorithm \cite{leisch2004flexmix}, the two trimmed likelihood approaches \cite{neykov2007robust, garcia2017robust},  the mixture bisquare \cite{bai2012robust}, mixture Laplacian\cite{song2014robust}, the mixture $t$ \cite{yao2014robust} approaches respectively. 

Note that for TLE and CWM, the choice of the trimming proportion, $\alpha$, needs to be pre-specified, and a large $\alpha$ will result in reduced efficiency, while a small $\alpha$ corrupt the parameter estimates. We always give the true ratio of outliers to TLE and CWM. In addition, a 5\% increase of the true outlier ratio is also given to CWM, as the authors recommended the use of a relatively larger trimming ratio than needed \cite{garcia2010robust} as a preventive procedure. We call the results with the true outlying ratio and 5\% larger ratio for CWM as CWM1 and CWM2 in Tables 1-4. 

We simulated data using two models with different number of covariates $P$, and number of components $K$.
To simulate outliers, a mean-shift parameter, $\gamma_{ij}$, is added to the mean structure for its observations in each mixture component. For each model, we considered scenarios with different error distributions and different levels of outlier contamination.\\

\noindent Model 1:
For each $i=1,...,N$, $y_i$ is independently generated with
% \begin{linenomath*}
\begin{equation*}
y_i = \left\{ 
\begin{array}{rcl} 
1-x_{i1}+x_{i2}+\gamma_{i1} + \epsilon_{i1} & \mbox{if}  & z_i=1 \\ 
1+3x_{i1}+x_{i2}+\gamma_{i2} + \epsilon_{i2} & \mbox{if} & z_i = 2
\end{array}\right.
\end{equation*}
% \end{linenomath*}
where $z_i$ is a component indicator generated from a Bernoulli distribution with $P(z_i=1)=0.43, P(z_i=1)=0.57$; $x_{i1}$ and $x_{i2}$ are independently generated from ${N}(0,1)$; and the error terms $\epsilon_{i1}$ and $\epsilon_{i2}$ have the same distribution as $\epsilon$. We consider the following five scenarios: \\

\noindent Scenario 1: $\epsilon \sim {N}(0,1), \gamma_{i1}=\gamma_{i2}=0 $, standard normal distribution. \\
Scenario 2: $\epsilon \sim {t}_1, \gamma_{i1}=\gamma_{i2}=0 $, $t$-distribution with degree of freedom of 1. \\
Scenario 3: $\epsilon \sim {t}_3, \gamma_{i1}=\gamma_{i2}=0 $, $t$-distribution with degree of freedom of 3. \\
Scenario 4: $\epsilon \sim {N}(0,1), P(\gamma_{i1}\in (4,6)) = P(\gamma_{i2}\in (4,6)) =0.05$, standard normal distribution with 5\% outlier contamination. \\
Scenario 5: $\epsilon \sim {N}(0,1), P(\gamma_{i1}\in (4,6)) = P(\gamma_{i2}\in (4,6)) =0.1$, standard normal distribution with 10\% outlier contamination. \\

\noindent Model 2:
For each $i=1,...,N$, $y_i$ is independently generated with
% \begin{linenomath*}
\begin{equation*}
y_i = \left\{ 
\begin{array}{rcl} 
1-x_{i1}+\gamma_{i1}  + \epsilon_{i1} & \mbox{if}  & z_i=1 \\ 
1+3x_{i1}+\gamma_{i2}  + \epsilon_{i2} & \mbox{if} &  z_i = 2 \\
-1+0.1x_{i1}+\gamma_{i3} + \epsilon_{i3} & \mbox{if} &  z_i = 3
\end{array}\right.
\end{equation*}
% \end{linenomath*}
where $z_i$ is a component indicator generated from a Multinomial distribution with $P(z_i=1)=0.3,P(z_i=2)=0.4,P(z_i=3)=0.3$. $x_{i1}$ is independently generated from ${N}(0,1)$; and the error terms $\epsilon_{i1}$, $\epsilon_{i2}$, $\epsilon_{i3}$ have the same distribution with $\epsilon$. \\

\noindent Scenario 1: $\epsilon \sim {N}(0,1), \gamma_{i1}=\gamma_{i2}=\gamma_{i3}=0 $, standard normal distribution. \\
Scenario 2: $\epsilon \sim {t}_1, \gamma_{i1}=\gamma_{i2}=\gamma_{i3}=0 $, $t$-distribution with degree of freedom of 1. \\
Scenario 3: $\epsilon \sim {t}_3, \gamma_{i1}=\gamma_{i2}=\gamma_{i3}=0 $, $t$-distribution with degree of freedom of 3. \\
Scenario 4: $\epsilon \sim {N}(0,1), P(\gamma_{i1}\in (4,6)) = P(\gamma_{i2}\in (4,6)) = P(\gamma_{i3}\in (4,6))=0.05$, standard normal distribution with 5\% outlier contamination. \\
Scenario 5: $\epsilon \sim {N}(0,1), P(\gamma_{i1}\in (4,6)) = P(\gamma_{i2}\in (4,6)) = P(\gamma_{i3}\in (4,6))=0.1$, standard normal distribution with 10\% outlier contamination. \\

For scenarios in both Models 1 and 2, we simulated data of sample sizes 200 and 400. The bias and mean square error (MSE) of the regression coefficients and mixing proportions are calculated for each competing methods over 100 repetitions, i.e., $\widehat{\operatorname{bias}}(\hat{\theta})=\frac{1}{N} \sum_{j=1}^{N} \hat{\theta}_{j}-\theta; \quad \widehat{\operatorname{MSE}}(\hat{\theta})=\frac{1}{N} \sum_{j=1}^{N}\left(\hat{\theta}_{j}-\theta\right)^{2}$.
The label switching issue \cite{celeux2000computational,stephens2000dealing,yao2009bayesian} creates some trouble on how to align the parameters of one component from predicted model to that of the true model. Different component orders in the predicted and true model might give totally different results and there are no widely accepted methods to adjust for that. In our simulation study, we simply choose to order the components in the estimated parameter matrix by minimizing the Euclidean distance to the true parameter matrix. \\

Table 1 reports the bias (MSE) of parameter estimates for the seven methods  under Model 1 with 200 simulated samples. It is a two component model with two independent variables. Note that in Scenarios 1,2,3, there are no added outliers, so the TLE estimates are the same as MLE estimates. For Scenarios 4 and 5, the true outlier proportions were given to TLE and CWM1, and 5\% increase of outlier proportion is given to CWM2. Similarly for Tables 2,3 and 4. When the component error terms are all normally distributed without outlier contamination (Scenario 1), all methods have comparable performances, except that the CWM methods perform much worse. In Scenario 2, where the error terms are $t$-distributed with degree of freedom of 1 (or Cauchy distribution), CAT resulted in the most favorable performances, followed by MIXBI and then CWM2. MLE, TLE and CWM1 estimates are severely corrupted, and MIXL and MIXT are also far off from the true values. For Scenario 3, where the error terms are $t$-distributed with degree of freedom of 3, CAT, MIXBI and MIXL all worked equally well in parameter estimation. The MLE, TLE, CWM1, CWM2 and MIXT estimates are slightly off from the true regression coefficients. For Scenarios 4 and 5, where the error terms are normally distributed, with 5\% and 10\% outlier contamination, CAT significantly outperformed all the other methods in terms of all the parameter estimations. Overall, CAT represents the most competitive one among the seven methods; TLE is very sensitive to heavy-tailed noise, and its performance is not ideal even given the true trimming proportion; for CWM, a larger than needed trimming proportion indeed leads to more accurate estimates.

Table 2 reports the bias (MSE) of the seven methods of parameter estimates under Model 1 with 400 simulated samples. It is a two component model with two independent variables. When the component error terms are all normally distributed without outlier contamination (Scenario 1), CAT performed comparatively good with all methods, except that the CWM methods performed much worse. In Scenario 2 and 3, where the error terms are $t$-distributed with degrees of freedom of 1 and 3, CAT performed the best among all, followed by MIXBI and then MIXL, while the rest of the methods seem to be severely corrupted by the heavy tail noise. Such corruption is much alleviated with the increase of degree of freedom, as $t$-distribution with higher degree of freedom is more like normal distribution. For Scenarios 4 and 5, where the error terms are normally distributed, with 5\% and 10\% outlier contamination, CAT significantly outperformed all the other methods in terms of parameter estimation. Again, CAT remains the best-performing one among the six methods.

Table 3 reports the bias (MSE) of the seven methods of parameter estimates under Model 2 with 200 simulated samples. It is a three component model with one independent variable. When the component error terms are all normally distributed without outlier contamination (Scenario 1), all six methods have comparable performances. In Scenario 2, where the error terms are $t$-distributed with degree of freedom of 1 (or Cauchy distribution), CAT and MIXBI resulted in comparable performances, with CAT being slightly more accurate and stable. MLE, TLE, CWM1, and MIXL estimates are severely biased, while CWM2 is highly unstable with large variance, and MIXT is also far off from the true values. For Scenario 3, where the error terms are $t$-distributed with degree of freedom of 3, CAT, MIXBI and MIXL all worked equally well in parameter estimation. The MLE, TLE, CWM and MIXT estimates are slightly off from the true regression coefficients. For Scenarios 4 and 5, where the error terms are normally distributed, with 5\% and 10\% outlier contamination, CAT significantly outperformed all the other methods in terms of parameter estimation. We show again the more competitive performance of CAT over others.

Table 4 reports the bias (MSE) of the six methods of parameter estimates under Model 2 with 400 simulated samples. It is a three component model with one independent variable. When the component error terms are all normally distributed without outlier contamination (Scenario 1), CAT and MIXBI both performed well, followed by MIXL; and they all outperformed the rest of the algorithms, even including the MLE estimates. We argue that even though no outliers or heavy tailed noise is added, there is still likely ``outlying" samples caused purely by chance; robust procedures like CAT and MIXBI work in preventive fashion to account for these random outliers, and are hence more robust even compared with the MLE. In Scenario 2, where the error terms are $t$-distributed with degree of freedom of 1 (or Cauchy distribution), CAT, CWM2, and MIXBI resulted in comparable performances, while the rest of the methods did much worse. For Scenario 3, where the error terms are $t$-distributed with degree of freedom of 3, CAT and MIXBI both worked equally well in parameter estimation, with CAT slightly better. The MLE, TLE, CWM, MIXL, and MIXT estimates are slightly off from the true regression coefficients. For Scenarios 4 and 5, where the error terms are normally distributed, with 5\% and 10\% outlier contamination, CAT significantly outperformed all the other methods in terms of parameter estimation. We demonstrated here that CAT is a more robust method compared with others.

Overall, CAT demonstrated its advantage over others with its strong capacity of adaptive trimming and robustness to both outliers and heavy-tailed noises. Of note, MIXBI is a robust algorithm whose performance is next to CAT, however, it conducts parameter estimations in presence of outliers, which may tend to bring down the accuracy in parameter estimation. For CWM, indeed a slightly large trimming ratio may lead to better estimates, but it seems to over-trim the data with its ``second" trimming step, which may be the reason for its low estimation efficiency. We also show that even when given the right outlier prevalence, TLE still can't produce results as robust as CAT, and it did worse when $P$ increases. In the case of no deliberate outlier contamination (Scenario 1), CAT performs better than or comparable to MLE, which is because CAT can automatically trim off observations that are highly noisy as a preventive procedure. When the error terms are $t$-distributed, CAT still remains the most robust method among all.

We also simulated data scenarios where the clusters are highly unbalanced, and basically, for $K=2$, we simulated data with cluster prior probability $P(z_i=1)=0.38,P(z_i=2)=0.62$; for $K=3$, we simulated data with cluster prior probability 
$P(z_i=1)=0.2,P(z_i=2)=0.32,P(z_i=3)=0.48$. We repeated the same experiments as shown in Tables 1-4 for the unbalanced cluster priors, and reported the findings in Supplementary Tables S1-S4 in Appendix. Similar to the balanced cases presented in Tables 1-4, CAT demonstrated the most desirable performances compared with all other methods.

\section{4 Real data application}
Colon adenocarcinoma is known as a heterogeneous disease with different molecular subtypes \cite{guinney2015consensus}. CREB3L1, or cyclic AMP responsive element-binding protein 3-like protein 1, is an important transcription factor that can suppress cell cycle \cite{denard2012doxorubicin, chen2013regulation}. The regulation of CREB3L1 is largely accomplished through epigenetic mechanisms in cancer and other disease \cite{chen2013regulation}. We now examine the latent relationship between CREB3L1 and one of its epigenetic regulators, cg16012690, in colon cancer. We collected the gene expression profile of CREB3L1 and the methylation profile of cg16012690 on  299 colon adenocarcinoma patients from the Cancer Genome Atlas (TCGA) cohort \cite{weinstein2013cancer}. 

We fitted the data using CAT, MLE, CWM, TLE, MIXBI, MIXL and MIXT with $K=2$, and the two regression lines are colored in red and shown in the top panel of Figure 1. We could see that even though the regression lines fitted by the methods are slightly different, they all seem to fit well the data points. In order to compare the robustness of the six methods, we added 10 high leverage points at $x=0$ (middle panel of Figure 1), and 10 high leverage points at $x=0.9$ (bottom panel of Figure 1), in both cases of which, $y$ is a random draw from uniform distribution $U(18,20)$. We then refitted the contaminated data using the seven methods for the two scenarios. For CWM and TLE, we give them both the true outlier proportion (CWM1 and TLE1), as well as 5\% increase of the true outlier proportion (CWM2 and TLE2). In the middle and bottom panels of Figure 1, the regression lines fitted using the intact data were shown as red lines, and those refitted using the contaminated data were shown as dashed blue lines. Clearly, CAT was robust to the added high leverage outliers in both scenarios, as the lines fitted before and after contamination overlap. For contaminated data at $x=0$ (middle panel), all methods except for CAT and MIXT were affected by the outliers as the regression lines after contamination tend to go through the outliers and were quite different from those before contamination. For contaminated data at $x=0.9$ (bottom panel), all other methods except for CAT, TLE and MIXBI were severely affected by the outliers, seen from a dramatic deviation of the regression lines before and after contamination. This shows that CAT achieves better performances over others in robustly revealing the latent relationship among genomic markers.

\begin{figure*} % picture
    \centering
    \includegraphics[width=16cm]{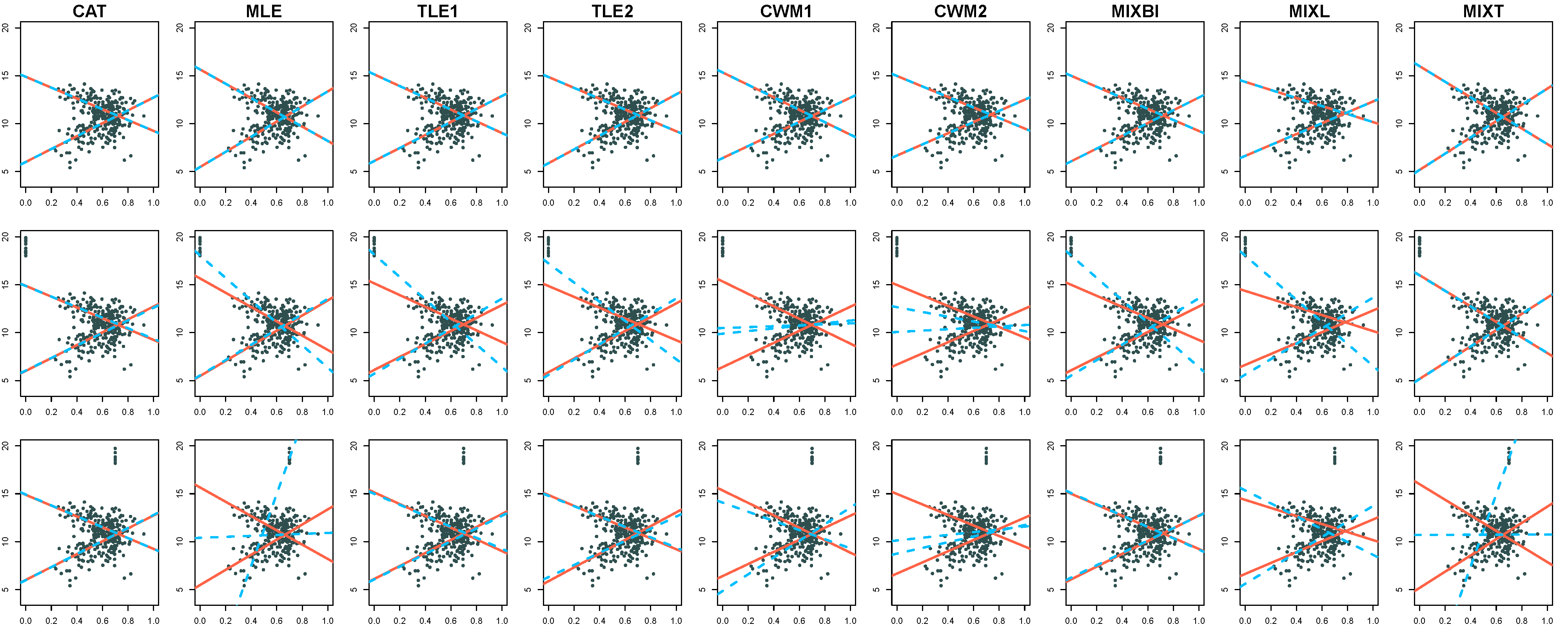}
    \caption{Mixture regression of CREB3L1 expression (y-axis) on cg16012690 methylation (x-axis) using different methods}
\end{figure*}

\section{Conclusion}
\noindent  We proposed a novel robust algorithm for solving FMGR using the Classification Expectation Maximization (CEM) algorithm, based on which, the outliers are more naturally defined, and robustness issue better and more conveniently handled. %it may bot be asymptotically consistent \cite{bryant1978asymptotic}. However,
Our method, CAT, enables the automatic detection of outliers and robust estimation of parameters simultaneously, which is not capable of by existing methods. The removal of outliers in final parameter estimation significantly increased the estimation efficiency than other algorithms; and the adoption of the highly robust least trimmed likelihood estimator within each component makes it possible to avoid the pre-specification of trimming parameter. Under the CEM framework, the adaptive trimming of a mixture model boils down into that of simple linear regression models corresponding to each component. Owing to its high breakdown point property, we assumed a high portion of outlying samples and the trimmed likelihood approach is optimized on only half of the samples within each component with the highest valued likelihood. This is the reason why CAT could be robust to outliers and heavy tailed noise. In summary, CAT is an robust mixture regression algorithm of high potential and practical utility in robustly mining the heterogeneous relations among noisy variables in the data booming era.

\section*{Declaration of conflicting interests}
The author(s) declared no potential conflicts of interest with respect to the research, authorship,and/or publication of this article.

\section*{Funding}
This project was partially supported by National Science Foundation (NSF IIS-1850360) and Indiana CTSI BERD Award and Showalter Trust Award to SC.

% Table 1
\begin{sidewaystable}
% \begin{table}
\captionsetup{font=normalsize}
\caption{$K$=2, $P$=2, $N$=200}
\centering
\small
\begin{tabular}{ccccccccc}
\hline
Scenario & CAT           & MLE              & TLE              & CWM1             & CWM2          & MIXBI         & MIXL           & MIXT           \\ \hline
\multicolumn{9}{c}{Scenario 1: $\epsilon \sim {N}(0,1)$ }  \\ \hline
1        & -0.030(0.130) & -0.030(0.120)    & -0.030(0.120)    & 0.060(0.450)     & 0.070(0.360)  & -0.030(0.130) & -0.080(0.210)  & -0.030(0.130)  \\
         & -0.010(0.130) & -0.010(0.120)    & -0.010(0.120)    & 0.010(0.310)     & -0.020(0.270) & -0.010(0.120) & -0.020(0.130)  & -0.010(0.130)  \\
         & 0.010(0.130)  & 0.010(0.120)     & 0.010(0.120)     & 0.350(0.720)     & 0.370(0.870)  & 0.010(0.120)  & 0.040(0.160)   & 0.020(0.120)   \\
         & -0.010(0.110) & 0.000(0.100)     & 0.000(0.100)     & -0.140(0.450)    & -0.190(0.450) & 0.000(0.100)  & -0.030(0.130)  & 0.010(0.100)   \\
         & 0.010(0.140)  & 0.010(0.130)     & 0.010(0.130)     & 0.030(0.250)     & -0.010(0.300) & 0.010(0.130)  & -0.010(0.170)  & 0.020(0.140)   \\
         & -0.020(0.110) & -0.030(0.110)    & -0.030(0.110)    & 0.010(0.230)     & 0.040(0.230)  & -0.020(0.110) & -0.020(0.140)  & -0.030(0.110)  \\
         & -0.000(0.050) & -0.000(0.040)    & -0.000(0.040)    & 0.040(0.140)     & 0.020(0.140)  & -0.000(0.040) & -0.010(0.040)  & -0.010(0.040)  \\
         & 0.000(0.050)  & 0.000(0.040)     & 0.000(0.040)     & -0.040(0.140)    & -0.020(0.140) & 0.000(0.040)  & 0.010(0.040)   & 0.010(0.040)   \\ \hline
\multicolumn{9}{c}{Scenario 2: $\epsilon \sim {t}_1$}    \\ \hline
2        & -0.030(0.820) & -30.780(283.320) & -30.780(283.320) & -39.680(438.250) & -0.260(2.480) & 0.030(0.280)  & 0.290(5.070)   & 8.300(52.430)  \\
         & 0.050(0.680)  & 36.360(302.980)  & 36.360(302.980)  & -1.130(30.160)   & -0.130(2.880) & -0.020(0.230) & 1.780(13.920)  & 0.290(16.510)  \\
         & 0.290(0.600)  & -32.720(152.040) & -32.720(152.040) & -39.430(262.650) & 1.200(1.650)  & 0.780(0.860)  & -1.490(11.580) & -9.540(35.410) \\
         & -0.030(0.290) & 73.300(323.140)  & 73.300(323.140)  & 9.680(30.910)    & -0.100(4.540) & 0.030(0.330)  & 3.110(14.440)  & 5.070(16.660)  \\
         & 0.140(1.480)  & -8.530(461.870)  & -8.530(461.870)  & -95.450(905.720) & -0.170(1.810) & 0.040(0.320)  & 0.280(10.860)  & 4.280(30.950)  \\
         & 0.010(0.310)  & -47.240(476.990) & -47.240(476.990) & -3.780(40.800)   & -0.700(5.450) & 0.010(0.260)  & -0.380(7.110)  & 3.450(18.460)  \\
         & 0.060(0.120)  & 0.100(0.420)     & 0.100(0.420)     & 0.010(0.410)     & 0.070(0.220)  & 0.170(0.180)  & 0.050(0.350)   & 0.020(0.410)   \\
         & -0.060(0.120) & -0.100(0.420)    & -0.100(0.420)    & -0.010(0.410)    & -0.070(0.220) & -0.170(0.180) & -0.050(0.350)  & -0.020(0.410)  \\ \hline
\multicolumn{9}{c}{Scenario 3: $\epsilon \sim {t}_3$}    \\ \hline
3        & -0.020(0.170) & -0.040(0.290)    & -0.040(0.290)    & 0.060(0.370)     & -0.070(0.430) & -0.020(0.170) & -0.040(0.180)  & -0.030(0.270)  \\
         & -0.020(0.140) & -0.030(0.180)    & -0.030(0.180)    & 0.010(0.260)     & 0.030(0.440)  & -0.020(0.150) & -0.030(0.160)  & -0.030(0.180)  \\
         & 0.020(0.160)  & 0.280(0.450)     & 0.280(0.450)     & 0.640(0.870)     & 0.560(0.870)  & 0.010(0.180)  & 0.040(0.180)   & 0.370(0.540)   \\
         & 0.000(0.150)  & 0.010(0.190)     & 0.010(0.190)     & -0.320(0.650)    & -0.220(0.550) & 0.030(0.140)  & -0.010(0.160)  & 0.020(0.190)   \\
         & -0.010(0.210) & -0.000(0.270)    & -0.000(0.270)    & 0.020(0.400)     & 0.020(0.490)  & -0.000(0.190) & -0.000(0.200)  & -0.010(0.260)  \\
         & -0.010(0.150) & 0.010(0.190)     & 0.010(0.190)     & 0.020(0.280)     & -0.030(0.290) & -0.000(0.150) & -0.030(0.170)  & 0.010(0.190)   \\
         & 0.020(0.060)  & 0.030(0.070)     & 0.030(0.070)     & 0.050(0.140)     & 0.040(0.150)  & 0.010(0.070)  & 0.010(0.060)   & 0.050(0.080)   \\
         & -0.020(0.060) & -0.030(0.070)    & -0.030(0.070)    & -0.050(0.140)    & -0.040(0.150) & -0.010(0.070) & -0.010(0.060)  & -0.050(0.080)  \\ \hline
\multicolumn{9}{c}{Scenario 4: 5\% added outliers}    \\ \hline
4        & -0.100(0.180) & -0.250(0.510)    & 0.220(1.500)     & 0.560(6.030)     & 0.480(3.520)  & -0.190(0.160) & -0.290(0.290)  & 0.200(0.440)   \\
         & -0.000(0.130) & 0.330(0.340)     & 0.270(1.010)     & 0.090(0.500)     & -0.160(1.540) & 0.060(0.130)  & 0.140(0.190)   & -0.010(0.210)  \\
         & -0.080(0.190) & 0.280(0.880)     & 1.780(2.400)     & 0.230(2.530)     & 0.530(1.190)  & -0.200(0.180) & -0.220(0.380)  & 0.940(0.630)   \\
         & 0.020(0.150)  & 0.040(0.770)     & -1.160(2.080)    & -0.190(0.820)    & -0.000(1.150) & 0.090(0.160)  & 0.180(0.180)   & -0.500(0.590)  \\
         & -0.110(0.240) & -0.440(1.020)    & -0.340(1.840)    & -0.140(1.300)    & -0.330(2.110) & -0.310(0.240) & -0.560(0.420)  & 0.520(0.710)   \\
         & 0.040(0.170)  & 0.710(0.640)     & 0.380(1.150)     & 0.120(0.770)     & 0.080(0.860)  & 0.140(0.170)  & 0.350(0.290)   & 0.060(0.400)   \\
         & 0.010(0.050)  & 0.000(0.150)     & 0.060(0.170)     & -0.010(0.160)    & 0.040(0.200)  & 0.010(0.050)  & -0.000(0.070)  & -0.060(0.150)  \\
         & -0.010(0.050) & -0.000(0.150)    & -0.060(0.170)    & 0.010(0.160)     & -0.040(0.200) & -0.010(0.050) & 0.000(0.070)   & 0.060(0.150)   \\ \hline
\multicolumn{9}{c}{Scenario 5: 10\% added outliers}    \\ \hline
5        & -0.000(0.140) & -0.120(0.450)    & -0.030(1.120)    & -0.030(0.210)    & 0.080(0.570)  & -0.030(0.140) & -0.100(0.170)  & 0.080(0.380)   \\
         & 0.020(0.120)  & 0.220(0.310)     & 0.020(1.140)     & -0.110(1.270)    & -0.140(0.590) & 0.040(0.110)  & 0.080(0.160)   & 0.050(0.150)   \\
         & -0.020(0.160) & 0.340(0.780)     & 1.990(2.660)     & 0.340(0.760)     & 0.570(0.910)  & -0.070(0.160) & -0.080(0.190)  & 0.500(0.630)   \\
         & 0.000(0.120)  & -0.050(0.580)    & -1.390(2.430)    & -0.110(0.430)    & -0.230(0.530) & 0.030(0.120)  & 0.060(0.120)   & -0.190(0.460)  \\
         & -0.040(0.150) & -0.240(0.740)    & -0.220(2.000)    & -0.010(0.340)    & -0.090(0.590) & -0.120(0.150) & -0.260(0.200)  & 0.130(0.670)   \\
         & 0.020(0.120)  & 0.410(0.670)     & -0.580(1.910)    & 0.150(1.130)     & 0.060(0.600)  & 0.040(0.120)  & 0.150(0.160)   & 0.040(0.200)   \\
         & 0.000(0.050)  & -0.030(0.160)    & 0.070(0.080)     & 0.030(0.140)     & 0.030(0.160)  & 0.010(0.050)  & 0.000(0.050)   & -0.040(0.120)  \\
         & -0.000(0.050) & 0.030(0.160)     & -0.070(0.080)    & -0.030(0.140)    & -0.030(0.160) & -0.010(0.050) & -0.000(0.050)  & 0.040(0.120)   \\ \hline
\end{tabular}
% \end{table}
\end{sidewaystable}

% Table 2
\begin{sidewaystable}
% \begin{table}[H]
\captionsetup{font=normalsize}
\caption{$K$=2, $P$=2,$N$=400}
\centering
\small
\begin{tabular}{ccccccccc}
\hline
Scenario & CAT           & MLE             & TLE             & CWM1            & CWM2          & MIXBI         & MIXL             & MIXT           \\ \hline
\multicolumn{9}{c}{Scenario 1: $\epsilon \sim {N}(0,1)$}                                                                                                                               \\ \hline
1        & -0.000(0.110) & 0.000(0.100)    & 0.000(0.100)    & 0.020(0.110)    & -0.000(0.380) & -0.000(0.100) & -0.000(0.130)    & 0.000(0.100)   \\
         & 0.000(0.090)  & 0.010(0.080)    & 0.010(0.080)    & 0.020(0.100)    & 0.010(0.110)  & 0.010(0.080)  & 0.010(0.100)     & 0.010(0.080)   \\
         & 0.000(0.080)  & 0.000(0.080)    & 0.000(0.080)    & 0.250(0.610)    & 0.290(0.570)  & 0.000(0.080)  & 0.020(0.100)     & 0.010(0.080)   \\
         & -0.020(0.080) & -0.010(0.070)   & -0.010(0.070)   & -0.080(0.310)   & -0.070(0.260) & -0.010(0.080) & -0.030(0.100)    & -0.000(0.090)  \\
         & 0.010(0.100)  & 0.000(0.080)    & 0.000(0.080)    & 0.020(0.110)    & 0.010(0.150)  & 0.000(0.090)  & -0.000(0.120)    & 0.000(0.080)   \\
         & -0.010(0.080) & -0.000(0.080)   & -0.000(0.080)   & 0.000(0.100)    & -0.000(0.110) & -0.010(0.080) & -0.010(0.090)    & -0.000(0.080)  \\
         & 0.000(0.030)  & 0.000(0.030)    & 0.000(0.030)    & 0.020(0.080)    & 0.020(0.100)  & 0.000(0.030)  & -0.000(0.030)    & 0.000(0.030)   \\
         & -0.000(0.030) & -0.000(0.030)   & -0.000(0.030)   & -0.020(0.080)   & -0.020(0.100) & -0.000(0.030) & 0.000(0.030)     & -0.000(0.030)  \\ \hline
         \multicolumn{9}{c}{Scenario 2: $\epsilon \sim {t}_1$}                                                                                                                               \\ \hline
2        & -0.020(0.220) & -0.710(43.750)  & -0.710(43.750)  & 1.630(44.370)   & -0.010(0.720) & 0.000(0.230)  & -7.310(84.040)   & 1.000(12.990)  \\
         & -0.010(0.130) & -4.470(73.010)  & -4.470(73.010)  & -9.330(92.790)  & -0.110(1.090) & 0.010(0.130)  & -20.000(207.170) & -0.190(15.170) \\
         & 0.330(0.450)  & -16.440(92.360) & -16.440(92.360) & -28.470(84.080) & 1.240(0.950)  & 0.930(0.650)  & -12.320(67.180)  & -2.880(13.170) \\
         & 0.000(0.160)  & 9.220(33.280)   & 9.220(33.280)   & 13.480(41.630)  & -0.620(1.330) & 0.060(0.160)  & 31.680(296.340)  & 6.010(13.120)  \\
         & 0.010(0.230)  & 3.840(67.430)   & 3.840(67.430)   & 0.260(45.000)   & 0.160(1.150)  & -0.010(0.210) & -10.200(83.010)  & 2.250(17.710)  \\
         & -0.020(0.130) & -2.420(63.500)  & -2.420(63.500)  & -4.210(61.080)  & -0.150(1.200) & -0.010(0.130) & 16.700(183.670)  & 0.290(18.170)  \\
         & 0.100(0.110)  & 0.050(0.430)    & 0.050(0.430)    & 0.070(0.440)    & 0.060(0.190)  & 0.220(0.120)  & 0.030(0.390)     & 0.090(0.420)   \\
         & -0.100(0.110) & -0.050(0.430)   & -0.050(0.430)   & -0.070(0.440)   & -0.060(0.190) & -0.220(0.120) & -0.030(0.390)    & -0.090(0.420)  \\ \hline
         \multicolumn{9}{c}{Scenario 3: $\epsilon \sim {t}_3$}                                                                                                                               \\ \hline
3        & -0.010(0.130) & 0.010(0.220)    & 0.010(0.220)    & 0.050(0.570)    & -0.000(0.270) & -0.000(0.130) & -0.010(0.140)    & 0.020(0.220)   \\
         & 0.010(0.110)  & 0.010(0.140)    & 0.010(0.140)    & -0.030(0.270)   & 0.080(0.820)  & 0.010(0.120)  & 0.020(0.120)     & 0.010(0.130)   \\
         & 0.010(0.120)  & 0.200(0.350)    & 0.200(0.350)    & 0.800(0.760)    & 0.570(0.850)  & 0.020(0.160)  & 0.020(0.130)     & 0.300(0.400)   \\
         & 0.010(0.100)  & 0.010(0.150)    & 0.010(0.150)    & -0.250(0.570)   & -0.110(0.770) & 0.030(0.090)  & 0.010(0.100)     & 0.030(0.130)   \\
         & -0.010(0.130) & -0.010(0.190)   & -0.010(0.190)   & 0.010(0.740)    & -0.030(0.210) & -0.010(0.120) & -0.010(0.120)    & -0.010(0.210)  \\
         & 0.000(0.110)  & -0.000(0.110)   & -0.000(0.110)   & -0.090(0.530)   & -0.110(0.660) & 0.000(0.100)  & 0.010(0.110)     & -0.000(0.120)  \\
         & 0.020(0.040)  & 0.020(0.050)    & 0.020(0.050)    & 0.060(0.160)    & 0.070(0.160)  & 0.020(0.050)  & 0.000(0.040)     & 0.040(0.060)   \\
         & -0.020(0.040) & -0.020(0.050)   & -0.020(0.050)   & -0.060(0.160)   & -0.070(0.160) & -0.020(0.050) & -0.000(0.040)    & -0.040(0.060)  \\ \hline
         \multicolumn{9}{c}{Scenario 4: 5\% added outliers}                                                                                                                               \\ \hline
4        & -0.050(0.120) & -0.100(0.480)   & -0.010(0.740)   & 0.390(2.590)    & 0.170(1.220)  & -0.170(0.120) & -0.280(0.150)    & 0.320(0.280)   \\
         & 0.010(0.090)  & 0.270(0.320)    & -0.010(0.560)   & -1.040(5.440)   & -0.290(2.220) & 0.070(0.090)  & 0.170(0.120)     & 0.030(0.120)   \\
         & -0.080(0.130) & 0.330(0.790)    & 2.160(1.730)    & 0.330(1.710)    & 0.100(1.720)  & -0.220(0.130) & -0.290(0.180)    & 1.010(0.400)   \\
         & 0.020(0.090)  & 0.040(0.650)    & -1.370(1.540)   & 0.300(2.290)    & 0.480(5.860)  & 0.090(0.100)  & 0.180(0.140)     & -0.500(0.550)  \\
         & -0.130(0.180) & -0.290(0.920)   & -0.130(1.140)   & -0.230(0.940)   & -0.000(0.720) & -0.350(0.170) & -0.590(0.200)    & 0.580(0.490)   \\
         & 0.010(0.100)  & 0.570(0.610)    & 0.030(1.130)    & 0.210(1.540)    & -0.400(4.380) & 0.130(0.120)  & 0.330(0.150)     & 0.030(0.180)   \\
         & 0.010(0.040)  & -0.010(0.120)   & 0.060(0.160)    & 0.050(0.200)    & 0.000(0.160)  & 0.000(0.040)  & -0.010(0.050)    & -0.080(0.130)  \\
         & -0.010(0.040) & 0.010(0.120)    & -0.060(0.160)   & -0.050(0.200)   & -0.000(0.160) & -0.000(0.040) & 0.010(0.050)     & 0.080(0.130)   \\ \hline
         \multicolumn{9}{c}{Scenario 5: 10\% added outliers}  \\ \hline
5        & -0.010(0.110) & -0.040(0.380)   & -0.160(1.290)   & -0.050(0.360)   & -0.030(0.250) & -0.050(0.110) & -0.110(0.130)    & 0.180(0.250)   \\
         & 0.010(0.090)  & 0.210(0.260)    & 0.010(1.220)    & -0.020(0.600)   & -0.010(0.350) & 0.030(0.090)  & 0.090(0.110)     & 0.040(0.160)   \\
         & -0.010(0.110) & 0.440(0.760)    & 1.480(2.270)    & 0.330(0.770)    & 0.280(0.710)  & -0.060(0.120) & -0.090(0.120)    & 0.730(0.590)   \\
         & 0.010(0.080)  & -0.070(0.610)   & -1.400(2.200)   & -0.090(0.440)   & -0.100(0.380) & 0.030(0.080)  & 0.070(0.100)     & -0.210(0.400)  \\
         & -0.010(0.120) & -0.130(0.700)   & 0.100(1.660)    & -0.030(0.310)   & -0.010(0.180) & -0.100(0.130) & -0.200(0.130)    & 0.370(0.430)   \\
         & 0.010(0.080)  & 0.360(0.450)    & 0.250(1.430)    & 0.050(0.500)    & 0.020(0.180)  & 0.050(0.090)  & 0.140(0.110)     & 0.040(0.310)   \\
         & 0.000(0.030)  & -0.020(0.130)   & 0.050(0.160)    & 0.030(0.120)    & 0.010(0.110)  & 0.000(0.030)  & -0.000(0.040)    & -0.020(0.120)  \\
         & -0.000(0.030) & 0.020(0.130)    & -0.050(0.160)   & -0.030(0.120)   & -0.010(0.110) & -0.000(0.030) & 0.000(0.040)     & 0.020(0.120)   \\ \hline
\end{tabular}
% \end{table}
\end{sidewaystable}

% Table 3
\begin{sidewaystable}
% \begin{table}[H]
\captionsetup{font=normalsize}
\caption{$K$=3, $P$=1,$N$=200}
\small
\begin{tabular}{ccccccccc}
\hline
Scenario & CAT           & MLE             & TLE             & CWM1             & CWM2          & MIXBI         & MIXL             & MIXT           \\ \hline
\multicolumn{9}{c}{Scenario 1: $\epsilon \sim {N}(0,1)$}   \\ \hline
1        & -0.040(0.290) & -0.080(0.430)   & -0.080(0.430)   & -0.330(0.510)    & -0.440(0.470) & -0.060(0.330) & -0.180(0.320)    & -0.170(0.430)  \\
         & -0.020(0.170) & 0.030(0.250)    & 0.030(0.250)    & 0.010(0.330)     & 0.030(0.440)  & -0.020(0.170) & 0.000(0.210)     & 0.040(0.220)   \\
         & -0.020(0.300) & 0.190(0.550)    & 0.190(0.550)    & 0.770(0.710)     & 0.860(0.720)  & 0.090(0.390)  & 0.270(0.410)     & 0.310(0.720)   \\
         & 0.040(0.330)  & 0.090(0.530)    & 0.090(0.530)    & 0.360(0.720)     & 0.360(0.880)  & -0.010(0.240) & -0.040(0.260)    & 0.040(0.430)   \\
         & 0.020(0.160)  & 0.030(0.160)    & 0.030(0.160)    & -0.100(0.400)    & -0.180(0.510) & 0.030(0.140)  & -0.000(0.170)    & 0.070(0.280)   \\
         & 0.030(0.270)  & 0.330(0.890)    & 0.330(0.890)    & 0.600(1.200)     & 0.740(1.160)  & 0.120(0.560)  & 0.170(0.380)     & 0.520(1.010)   \\
         & 0.010(0.080)  & 0.030(0.100)    & 0.030(0.100)    & 0.060(0.140)     & 0.050(0.140)  & 0.010(0.100)  & 0.010(0.070)     & 0.050(0.120)   \\
         & -0.010(0.060) & -0.030(0.080)   & -0.030(0.080)   & -0.040(0.110)    & -0.060(0.120) & -0.010(0.070) & -0.000(0.060)    & -0.030(0.090)  \\
         & -0.000(0.090) & -0.000(0.080)   & -0.000(0.080)   & -0.010(0.130)    & 0.000(0.130)  & -0.000(0.100) & -0.000(0.090)    & -0.020(0.110)  \\ \hline
         \multicolumn{9}{c}{Scenario 2: $\epsilon \sim {t}_1$}    \\ \hline
2        & 0.080(1.380)  & 7.190(36.260)   & 7.190(36.260)   & 7.140(75.550)    & -0.360(1.470) & -0.240(0.740) & 14.400(92.190)   & -0.230(8.140)  \\
         & 0.140(0.590)  & 10.580(127.780) & 10.580(127.780) & 13.280(66.830)   & -0.190(1.530) & -0.070(0.350) & 29.740(172.710)  & 2.110(14.780)  \\
         & 0.500(1.030)  & -22.530(74.610) & -22.530(74.610) & -12.070(49.140)  & 0.100(3.500)  & 0.450(2.540)  & -41.960(132.220) & -1.460(5.950)  \\
         & 0.360(0.960)  & -9.090(27.760)  & -9.090(27.760)  & -43.040(188.940) & 0.240(2.750)  & 0.570(0.890)  & -39.440(208.580) & -4.210(12.880) \\
         & -0.010(0.460) & 48.320(145.340) & 48.320(145.340) & 21.960(98.690)   & -0.190(1.670) & -0.190(0.640) & 42.380(142.150)  & 6.830(18.380)  \\
         & 0.480(1.280)  & -12.550(82.170) & -12.550(82.170) & -2.500(29.920)   & 0.050(3.620)  & 0.540(1.080)  & -27.790(142.750) & 0.690(3.730)   \\
         & 0.040(0.130)  & 0.060(0.270)    & 0.060(0.270)    & 0.030(0.250)     & 0.060(0.160)  & 0.050(0.190)  & 0.080(0.310)     & 0.020(0.210)   \\
         & -0.090(0.090) & -0.110(0.260)   & -0.110(0.260)   & -0.120(0.230)    & -0.090(0.160) & -0.090(0.170) & -0.130(0.270)    & -0.080(0.180)  \\
         & 0.050(0.130)  & 0.050(0.260)    & 0.050(0.260)    & 0.090(0.270)     & 0.030(0.170)  & 0.040(0.190)  & 0.050(0.290)     & 0.060(0.200)   \\ \hline
         \multicolumn{9}{c}{Scenario 3: $\epsilon \sim {t}_3$}   \\ \hline
3        & 0.040(0.530)  & 0.130(1.600)    & 0.130(1.600)    & -0.120(0.900)    & -0.390(0.640) & -0.170(0.470) & -0.260(0.450)    & -0.250(0.590)  \\
         & 0.070(0.390)  & 0.310(2.620)    & 0.310(2.620)    & -0.090(0.480)    & 0.050(0.420)  & 0.040(0.240)  & 0.030(0.290)     & 0.070(0.300)   \\
         & 0.130(0.600)  & 0.520(1.580)    & 0.520(1.580)    & 0.720(0.890)     & 0.780(0.650)  & 0.340(0.670)  & 0.400(0.660)     & 0.570(0.760)   \\
         & 0.110(0.390)  & 0.120(1.520)    & 0.120(1.520)    & 0.480(1.080)     & 0.330(0.740)  & 0.020(0.360)  & -0.020(0.380)    & 0.220(0.770)   \\
         & 0.000(0.190)  & 0.460(1.650)    & 0.460(1.650)    & -0.230(0.590)    & -0.220(0.460) & 0.070(0.230)  & 0.010(0.240)     & 0.050(0.330)   \\
         & 0.200(0.780)  & 0.310(2.030)    & 0.310(2.030)    & 0.500(1.190)     & 0.490(1.010)  & 0.340(0.880)  & 0.190(0.810)     & 0.550(1.310)   \\
         & 0.010(0.100)  & 0.010(0.190)    & 0.010(0.190)    & 0.030(0.150)     & 0.020(0.150)  & 0.010(0.150)  & 0.010(0.110)     & 0.030(0.140)   \\
         & -0.030(0.080) & -0.070(0.130)   & -0.070(0.130)   & -0.050(0.130)    & -0.050(0.120) & -0.040(0.090) & -0.010(0.080)    & -0.060(0.110)  \\
         & 0.030(0.110)  & 0.060(0.170)    & 0.060(0.170)    & 0.020(0.140)     & 0.030(0.150)  & 0.040(0.160)  & 0.000(0.110)     & 0.030(0.140)   \\ \hline
         \multicolumn{9}{c}{Scenario 4: 5\% added outliers}                                                                                                                               \\ \hline
4        & -0.420(0.480) & -1.300(0.890)   & -0.630(1.590)   & 0.180(4.760)     & -0.470(1.010) & -0.910(0.630) & -1.150(0.840)    & -0.810(0.450)  \\
         & 0.110(0.420)  & 0.720(0.770)    & -0.450(1.530)   & -0.270(1.360)    & -0.110(0.710) & 0.260(0.500)  & 0.380(0.560)     & 0.140(0.330)   \\
         & 0.300(0.590)  & 0.190(1.690)    & 0.980(1.370)    & 0.880(0.760)     & 0.890(0.690)  & 0.510(1.070)  & 0.290(1.370)     & 1.160(0.650)   \\
         & -0.200(0.550) & -0.430(1.730)   & 1.520(2.690)    & -0.440(3.050)    & 0.250(1.020)  & -0.330(1.200) & -0.650(1.360)    & 0.360(0.730)   \\
         & 0.180(0.560)  & 0.890(0.970)    & -0.910(2.240)   & 0.170(1.140)     & 0.010(0.810)  & 0.450(0.670)  & 0.480(0.610)     & 0.170(0.380)   \\
         & 0.260(0.900)  & -0.780(2.410)   & 0.900(2.280)    & 0.390(1.480)     & 0.630(1.140)  & 0.340(1.750)  & -0.670(2.650)    & 0.430(1.290)   \\
         & 0.030(0.120)  & 0.110(0.210)    & 0.010(0.160)    & 0.020(0.170)     & 0.050(0.130)  & 0.120(0.140)  & 0.050(0.170)     & 0.120(0.170)   \\
         & -0.010(0.100) & -0.090(0.120)   & -0.020(0.160)   & -0.070(0.140)    & -0.070(0.140) & -0.040(0.100) & -0.030(0.140)    & -0.040(0.090)  \\
         & -0.020(0.100) & -0.020(0.250)   & 0.010(0.150)    & 0.050(0.150)     & 0.020(0.160)  & -0.080(0.130) & -0.020(0.160)    & -0.080(0.180)  \\ \hline
         \multicolumn{9}{c}{Scenario 5: 10\% added outliers}                                                                                                                               \\ \hline
5        & -0.140(0.410) & -0.810(0.950)   & -0.620(1.270)   & 0.720(8.710)     & -0.270(1.140) & -0.440(0.370) & -0.770(0.590)    & -0.600(0.610)  \\
         & 0.030(0.170)  & 0.530(0.830)    & -0.870(1.160)   & -0.030(0.410)    & -0.060(0.560) & 0.090(0.270)  & 0.390(0.580)     & 0.190(0.530)   \\
         & 0.090(0.420)  & 0.390(1.620)    & 0.790(1.350)    & 0.700(0.970)     & 0.670(0.660)  & 0.340(0.600)  & 0.590(1.240)     & 0.890(1.110)   \\
         & -0.160(0.390) & 0.330(1.570)    & 0.930(2.040)    & -0.420(5.000)    & 0.400(1.080)  & -0.100(0.360) & -0.140(0.930)    & 0.550(1.050)   \\
         & 0.030(0.250)  & 0.620(1.090)    & -2.060(2.550)   & 0.000(0.440)     & -0.120(0.530) & 0.150(0.460)  & 0.430(0.680)     & 0.190(0.540)   \\
         & 0.090(0.520)  & -0.590(2.560)   & 0.350(2.110)    & 0.480(1.290)     & 0.400(1.200)  & 0.270(0.770)  & 0.080(2.460)     & -0.020(2.020)  \\
         & -0.000(0.090) & 0.110(0.230)    & -0.010(0.180)   & 0.050(0.130)     & 0.040(0.150)  & 0.060(0.130)  & 0.090(0.150)     & 0.060(0.220)   \\
         & 0.000(0.060)  & -0.080(0.150)   & -0.000(0.200)   & -0.060(0.110)    & -0.050(0.130) & -0.010(0.090) & -0.070(0.170)    & -0.040(0.130)  \\
         & -0.000(0.100) & -0.020(0.250)   & 0.010(0.170)    & 0.000(0.120)     & 0.010(0.140)  & -0.060(0.120) & -0.020(0.150)    & -0.010(0.240)  \\ \hline
\end{tabular}
% \end{table}
\end{sidewaystable}

% Table 4
\begin{sidewaystable}
% \begin{table}[H]
\captionsetup{font=normalsize}
\caption{$K$=3, $P$=1,$N$=400}
\centering
\small
\begin{tabular}{ccccccccc}
\hline
Scenario & CAT           & MLE              & TLE              & CWM1               & CWM2          & MIXBI         & MIXL              & MIXT           \\ \hline
\multicolumn{9}{c}{Scenario 1: $\epsilon \sim {N}(0,1)$}    \\ \hline
1        & -0.000(0.220) & -0.080(0.320)    & -0.080(0.320)    & -0.310(0.580)      & -0.370(0.500) & -0.030(0.200) & -0.140(0.230)     & -0.080(0.250)  \\
         & 0.010(0.170)  & 0.020(0.180)     & 0.020(0.180)     & -0.040(0.360)      & -0.050(0.400) & -0.000(0.110) & -0.000(0.130)     & 0.000(0.110)   \\
         & 0.040(0.360)  & 0.160(0.510)     & 0.160(0.510)     & 0.900(0.660)       & 0.800(0.730)  & 0.010(0.200)  & 0.160(0.250)      & 0.070(0.310)   \\
         & 0.090(0.180)  & 0.100(0.450)     & 0.100(0.450)     & 0.530(0.780)       & 0.530(0.870)  & 0.010(0.150)  & -0.030(0.190)     & -0.020(0.170)  \\
         & 0.010(0.110)  & 0.020(0.120)     & 0.020(0.120)     & -0.040(0.390)      & -0.090(0.510) & 0.020(0.090)  & 0.010(0.110)      & 0.020(0.130)   \\
         & -0.020(0.460) & 0.210(0.760)     & 0.210(0.760)     & 0.630(1.130)       & 0.700(1.180)  & -0.030(0.180) & 0.070(0.210)      & 0.090(0.390)   \\
         & 0.010(0.070)  & 0.020(0.100)     & 0.020(0.100)     & 0.060(0.150)       & 0.080(0.150)  & 0.000(0.060)  & 0.000(0.050)      & 0.010(0.070)   \\
         & -0.010(0.060) & -0.020(0.060)    & -0.020(0.060)    & -0.060(0.110)      & -0.070(0.130) & 0.000(0.040)  & 0.000(0.040)      & -0.000(0.050)  \\
         & 0.000(0.070)  & -0.000(0.070)    & -0.000(0.070)    & 0.000(0.110)       & -0.010(0.110) & -0.000(0.070) & -0.000(0.060)     & -0.010(0.070)  \\ \hline
 \multicolumn{9}{c}{Scenario 2: $\epsilon \sim {t}_1$}    \\ \hline
2        & -0.060(1.060) & 52.300(421.560)  & 52.300(421.560)  & 210.880(2052.960)  & 0.420(3.520)  & -0.400(0.500) & 215.860(1933.390) & 2.760(12.470)  \\
         & 0.090(0.510)  & 14.380(125.220)  & 14.380(125.220)  & -1.850(69.520)     & 0.170(1.950)  & -0.030(0.300) & 192.110(1939.830) & 0.430(6.240)   \\
         & 0.850(0.750)  & -37.930(147.630) & -37.930(147.630) & -21.000(71.890)    & 0.670(1.300)  & 0.920(0.670)  & -66.660(389.220)  & -1.250(5.190)  \\
         & 0.500(0.920)  & -81.630(437.820) & -81.630(437.820) & -435.630(4236.130) & 0.150(3.780)  & 0.520(0.870)  & 291.150(3140.910) & -4.990(13.140) \\
         & 0.020(0.350)  & 74.650(183.210)  & 74.650(183.210)  & 38.330(147.010)    & -0.440(0.970) & -0.100(0.410) & 390.210(3137.850) & 3.780(14.810)  \\
         & 0.660(1.130)  & -13.880(108.340) & -13.880(108.340) & -0.010(30.140)     & 0.510(1.260)  & 0.580(0.860)  & 12.430(223.510)   & -0.460(5.600)  \\
         & 0.030(0.150)  & 0.070(0.320)     & 0.070(0.320)     & 0.030(0.260)       & 0.040(0.170)  & 0.030(0.160)  & 0.020(0.310)      & 0.060(0.230)   \\
         & -0.090(0.100) & -0.140(0.270)    & -0.140(0.270)    & -0.070(0.300)      & -0.080(0.150) & -0.110(0.180) & -0.120(0.300)     & -0.100(0.180)  \\
         & 0.060(0.150)  & 0.070(0.310)     & 0.070(0.310)     & 0.040(0.270)       & 0.040(0.160)  & 0.080(0.200)  & 0.100(0.340)      & 0.040(0.220)   \\ \hline
         \multicolumn{9}{c}{Scenario 3: $\epsilon \sim {t}_3$}    \\ \hline
3        & 0.100(0.320)  & -0.450(1.120)    & -0.450(1.120)    & -0.420(0.740)      & -0.340(0.560) & -0.110(0.370) & -0.160(0.310)     & -0.600(0.400)  \\
         & -0.040(0.160) & 0.140(1.180)     & 0.140(1.180)     & -0.160(0.970)      & 0.020(0.370)  & -0.030(0.170) & -0.030(0.170)     & -0.020(0.470)  \\
         & -0.000(0.250) & -1.620(20.930)   & -1.620(20.930)   & 0.860(1.100)       & 0.690(0.670)  & 0.130(0.450)  & -1.840(20.780)    & 1.000(0.790)   \\
         & 0.150(0.430)  & 0.200(2.030)     & 0.200(2.030)     & 0.570(0.740)       & 0.600(0.780)  & 0.030(0.330)  & -0.030(0.210)     & 0.430(0.600)   \\
         & -0.010(0.130) & 0.250(1.400)     & 0.250(1.400)     & -0.190(1.330)      & -0.220(0.420) & 0.030(0.130)  & 0.010(0.170)      & 0.020(0.500)   \\
         & 0.010(0.360)  & -2.110(28.490)   & -2.110(28.490)   & 0.610(1.250)       & 0.490(1.060)  & 0.150(0.490)  & -2.680(28.370)    & 1.240(1.220)   \\
         & 0.010(0.070)  & 0.090(0.190)     & 0.090(0.190)     & 0.040(0.150)       & 0.070(0.150)  & 0.030(0.120)  & 0.010(0.080)      & 0.140(0.140)   \\
         & -0.020(0.060) & -0.090(0.130)    & -0.090(0.130)    & -0.060(0.130)      & -0.070(0.120) & -0.030(0.080) & -0.010(0.060)     & -0.090(0.120)  \\
         & 0.020(0.070)  & -0.010(0.180)    & -0.010(0.180)    & 0.020(0.150)       & 0.010(0.140)  & -0.000(0.140) & -0.000(0.080)     & -0.050(0.130)  \\ \hline
         \multicolumn{9}{c}{Scenario 4: 5\% added outliers}       \\ \hline
4        & -0.400(0.490) & -1.280(0.940)    & -0.540(0.920)    & 0.310(4.980)       & -0.510(0.420) & -0.850(0.570) & -1.120(0.670)     & -0.770(0.420)  \\
         & 0.070(0.380)  & 0.620(0.640)     & -0.370(1.010)    & -0.380(1.010)      & -0.110(0.420) & 0.250(0.470)  & 0.570(0.630)      & 0.120(0.240)   \\
         & 0.270(0.470)  & 0.460(1.670)     & 1.060(1.240)     & 0.990(0.590)       & 0.880(0.780)  & 0.500(0.820)  & 0.580(1.500)      & 1.410(0.600)   \\
         & -0.330(0.630) & -0.530(1.960)    & 1.950(1.760)     & -0.430(3.360)      & 0.460(0.770)  & -0.370(1.130) & -0.520(1.200)     & 0.440(0.730)   \\
         & 0.110(0.400)  & 0.770(0.790)     & -1.210(1.700)    & 0.240(1.010)       & -0.080(0.460) & 0.380(0.570)  & 0.680(0.690)      & 0.130(0.210)   \\
         & 0.140(0.560)  & -0.740(2.460)    & 0.910(1.960)     & 0.550(1.370)       & 0.690(1.260)  & 0.330(1.280)  & -0.200(2.920)     & 0.670(1.120)   \\
         & 0.020(0.090)  & 0.100(0.220)     & 0.030(0.140)     & 0.050(0.150)       & 0.080(0.140)  & 0.110(0.120)  & 0.080(0.180)      & 0.150(0.150)   \\
         & -0.010(0.060) & -0.070(0.100)    & -0.030(0.160)    & -0.060(0.130)      & -0.070(0.130) & -0.010(0.090) & -0.070(0.160)     & -0.030(0.070)  \\
         & -0.010(0.080) & -0.030(0.230)    & 0.000(0.150)     & 0.010(0.140)       & -0.010(0.130) & -0.100(0.110) & -0.010(0.150)     & -0.110(0.160)  \\ \hline
         \multicolumn{9}{c}{Scenario 5: 10\% added outliers}      \\ \hline
5        & -0.060(0.270) & -1.060(0.950)    & -0.660(1.410)    & 0.100(5.540)       & -0.350(0.570) & -0.530(0.380) & -0.870(0.600)     & -0.670(0.640)  \\
         & 0.030(0.160)  & 0.320(0.480)     & -0.270(1.340)    & -0.070(0.380)      & -0.090(0.330) & 0.050(0.150)  & 0.260(0.460)      & 0.130(0.420)   \\
         & 0.040(0.260)  & 0.580(1.480)     & 1.330(1.260)     & 0.850(0.600)       & 0.770(0.630)  & 0.310(0.530)  & 0.560(0.900)      & 1.270(0.550)   \\
         & 0.010(0.210)  & -0.220(1.970)    & 1.080(2.750)     & -0.290(3.430)      & 0.370(0.790)  & -0.140(0.710) & -0.400(0.930)     & 0.610(0.920)   \\
         & 0.020(0.120)  & 0.310(0.420)     & -1.390(2.090)    & 0.000(0.440)       & -0.070(0.480) & 0.110(0.250)  & 0.320(0.700)      & 0.160(0.430)   \\
         & -0.030(0.370) & -0.660(2.260)    & 0.740(2.180)     & 0.470(1.200)       & 0.480(1.040)  & 0.170(0.600)  & -0.060(1.820)     & 0.400(1.220)   \\
         & 0.000(0.060)  & 0.090(0.250)     & 0.040(0.160)     & 0.050(0.140)       & 0.040(0.150)  & 0.060(0.120)  & 0.030(0.160)      & 0.100(0.220)   \\
         & -0.000(0.040) & -0.040(0.110)    & -0.010(0.170)    & -0.030(0.120)      & -0.040(0.130) & 0.010(0.060)  & -0.030(0.140)     & -0.030(0.100)  \\
         & -0.000(0.070) & -0.050(0.250)    & -0.030(0.180)    & -0.020(0.120)      & 0.010(0.140)  & -0.070(0.120) & -0.000(0.140)     & -0.070(0.220)  \\         \hline
\end{tabular}
% \end{table}
\end{sidewaystable}

\bibliographystyle{unsrt}  
\bibliography{ref.bib}

\begin{thebibliography}{10}

\bibitem{goldfeld1973estimation}
Stephen Goldfeld and Richard Quandt.
\newblock The estimation of structural shifts by switching regressions.
\newblock In {\em Annals of Economic and Social Measurement, Volume 2, number
  4}, pages 475--485. NBER, 1973.

\bibitem{bohning1999computer}
Dankmar B{\"o}hning.
\newblock {\em Computer-assisted analysis of mixtures and applications:
  meta-analysis, disease mapping and others}, volume~81.
\newblock CRC press, 1999.

\bibitem{hennig2000identifiablity}
Christian Hennig.
\newblock Identifiablity of models for clusterwise linear regression.
\newblock {\em Journal of Classification}, 17(2), 2000.

\bibitem{jiang1999hierarchical}
Wenxin Jiang and Martin~A Tanner.
\newblock Hierarchical mixtures-of-experts for exponential family regression
  models: approximation and maximum likelihood estimation.
\newblock {\em Annals of Statistics}, pages 987--1011, 1999.

\bibitem{mclachlan2004finite}
Geoffrey~J McLachlan and David Peel.
\newblock {\em Finite mixture models}.
\newblock John Wiley \& Sons, 2004.

\bibitem{xu1996convergence}
Lei Xu and Michael~I Jordan.
\newblock On convergence properties of the em algorithm for gaussian mixtures.
\newblock {\em Neural computation}, 8(1):129--151, 1996.

\bibitem{fruhwirth2006finite}
Sylvia Fr{\"u}hwirth-Schnatter.
\newblock {\em Finite mixture and Markov switching models}.
\newblock Springer Science \& Business Media, 2006.

\bibitem{yu2020selective}
Chun Yu, Weixin Yao, and Guangren Yang.
\newblock A selective overview and comparison of robust mixture regression
  estimators.
\newblock {\em International Statistical Review}, 88(1):176--202, 2020.

\bibitem{markatou2000mixture}
Marianthi Markatou.
\newblock Mixture models, robustness, and the weighted likelihood methodology.
\newblock {\em Biometrics}, 56(2):483--486, 2000.

\bibitem{shen2004outlier}
Hong-bin Shen, Jie Yang, and Shi-tong Wang.
\newblock Outlier detecting in fuzzy switching regression models.
\newblock In {\em International Conference on Artificial Intelligence:
  Methodology, Systems, and Applications}, pages 208--215. Springer, 2004.

\bibitem{bai2012robust}
Xiuqin Bai, Weixin Yao, and John~E Boyer.
\newblock Robust fitting of mixture regression models.
\newblock {\em Computational Statistics \& Data Analysis}, 56(7):2347--2359,
  2012.

\bibitem{bashir2012robust}
Shaheena Bashir and EM~Carter.
\newblock Robust mixture of linear regression models.
\newblock {\em Communications in Statistics-Theory and Methods},
  41(18):3371--3388, 2012.

\bibitem{song2014robust}
Weixing Song, Weixin Yao, and Yanru Xing.
\newblock Robust mixture regression model fitting by laplace distribution.
\newblock {\em Computational Statistics \& Data Analysis}, 71:128--137, 2014.

\bibitem{yao2014robust}
Weixin Yao, Yan Wei, and Chun Yu.
\newblock Robust mixture regression using the t-distribution.
\newblock {\em Computational Statistics \& Data Analysis}, 71:116--127, 2014.

\bibitem{peel2000robust}
David Peel and Geoffrey~J McLachlan.
\newblock Robust mixture modelling using the t distribution.
\newblock {\em Statistics and computing}, 10(4):339--348, 2000.

\bibitem{yu2017new}
Chun Yu, Weixin Yao, and Kun Chen.
\newblock A new method for robust mixture regression.
\newblock {\em Canadian Journal of Statistics}, 45(1):77--94, 2017.

\bibitem{neykov2007robust}
Neyko Neykov, Peter Filzmoser, R~Dimova, and Plamen Neytchev.
\newblock Robust fitting of mixtures using the trimmed likelihood estimator.
\newblock {\em Computational Statistics \& Data Analysis}, 52(1):299--308,
  2007.

\bibitem{celeux1992classification}
Gilles Celeux and G{\'e}rard Govaert.
\newblock A classification em algorithm for clustering and two stochastic
  versions.
\newblock {\em Computational statistics \& Data analysis}, 14(3):315--332,
  1992.

\bibitem{dougru2018robust}
Fatma~Zehra Do{\u{g}}ru and Olcay Arslan.
\newblock Robust mixture regression modeling using the least trimmed squares
  (lts)-estimation method.
\newblock {\em Communications in Statistics-Simulation and Computation},
  47(7):2184--2196, 2018.

\bibitem{rousseeuw1984least}
Peter~J Rousseeuw.
\newblock Least median of squares regression.
\newblock {\em Journal of the American statistical association},
  79(388):871--880, 1984.

\bibitem{garcia2010robust}
Luis~Angel Garc{\'\i}a-Escudero, Alfonso Gordaliza, Agust{\'\i}n
  Mayo-{\'I}scar, and Roberto San~Mart{\'\i}n.
\newblock Robust clusterwise linear regression through trimming.
\newblock {\em Computational Statistics \& Data Analysis}, 54(12):3057--3069,
  2010.

\bibitem{garcia2017robust}
Luis~Angel Garc{\'\i}a-Escudero, Alfonso Gordaliza, Francesca Greselin,
  Salvatore Ingrassia, and Agust{\'\i}n Mayo-{\'I}scar.
\newblock Robust estimation of mixtures of regressions with random covariates,
  via trimming and constraints.
\newblock {\em Statistics and Computing}, 27(2):377--402, 2017.

\bibitem{guinney2015consensus}
Justin Guinney, Rodrigo Dienstmann, Xin Wang, Aur{\'e}lien De~Reyni{\`e}s,
  Andreas Schlicker, Charlotte Soneson, Laetitia Marisa, Paul Roepman, Gift
  Nyamundanda, Paolo Angelino, et~al.
\newblock The consensus molecular subtypes of colorectal cancer.
\newblock {\em Nature medicine}, 21(11):1350--1356, 2015.

\bibitem{faria2010fitting}
Susana Faria and Gilda Soromenho.
\newblock Fitting mixtures of linear regressions.
\newblock {\em Journal of Statistical Computation and Simulation},
  80(2):201--225, 2010.

\bibitem{siegel1982robust}
Andrew~F Siegel.
\newblock Robust regression using repeated medians.
\newblock {\em Biometrika}, 69(1):242--244, 1982.

\bibitem{blomer2016hard}
Johannes Bl{\"o}mer, Sascha Brauer, and Kathrin Bujna.
\newblock Hard-clustering with gaussian mixture models.
\newblock {\em arXiv preprint arXiv:1603.06478}, 2016.

\bibitem{donoho1983notion}
David~L Donoho and Peter~J Huber.
\newblock The notion of breakdown point.
\newblock {\em A festschrift for Erich L. Lehmann}, 157184, 1983.

\bibitem{yu2017robust}
Chun Yu and Weixin Yao.
\newblock Robust linear regression: A review and comparison.
\newblock {\em Communications in Statistics-Simulation and Computation},
  46(8):6261--6282, 2017.

\bibitem{rousseeuw2005robust}
Peter~J Rousseeuw and Annick~M Leroy.
\newblock {\em Robust regression and outlier detection}, volume 589.
\newblock John wiley \& sons, 2005.

\bibitem{rousseeuw1984robust}
Peter Rousseeuw and Victor Yohai.
\newblock Robust regression by means of s-estimators.
\newblock In {\em Robust and nonlinear time series analysis}, pages 256--272.
  Springer, 1984.

\bibitem{pison2002small}
Greet Pison, Stefan Van~Aelst, and G~Willems.
\newblock Small sample corrections for lts and mcd.
\newblock {\em Metrika}, 55(1-2):111--123, 2002.

\bibitem{leroy1987robust}
Annick~M Leroy and Peter~J Rousseeuw.
\newblock Robust regression and outlier detection.
\newblock {\em Wiley Series in Probability and Mathematical Statistics, New
  York: Wiley, 1987}, 1987.

\bibitem{rousseeuw1999fast}
Peter~J Rousseeuw and Katrien~Van Driessen.
\newblock A fast algorithm for the minimum covariance determinant estimator.
\newblock {\em Technometrics}, 41(3):212--223, 1999.

\bibitem{leisch2004flexmix}
Friedrich Leisch.
\newblock Flexmix: A general framework for finite mixture models and latent
  glass regression in r.
\newblock {\em Journal of Statistical Software}, 11(8):1--18, 2004.

\bibitem{celeux2000computational}
Gilles Celeux, Merrilee Hurn, and Christian~P Robert.
\newblock Computational and inferential difficulties with mixture posterior
  distributions.
\newblock {\em Journal of the American Statistical Association},
  95(451):957--970, 2000.

\bibitem{stephens2000dealing}
Matthew Stephens.
\newblock Dealing with label switching in mixture models.
\newblock {\em Journal of the Royal Statistical Society: Series B (Statistical
  Methodology)}, 62(4):795--809, 2000.

\bibitem{yao2009bayesian}
Weixin Yao and Bruce~G Lindsay.
\newblock Bayesian mixture labeling by highest posterior density.
\newblock {\em Journal of the American Statistical Association},
  104(486):758--767, 2009.

\bibitem{denard2012doxorubicin}
Bray Denard, Ching Lee, and Jin Ye.
\newblock Doxorubicin blocks proliferation of cancer cells through proteolytic
  activation of creb3l1.
\newblock {\em Elife}, 1:e00090, 2012.

\bibitem{chen2013regulation}
Qiuyue Chen.
\newblock {\em Regulation of Expression and Regulated Intramembrane Proteolysis
  of CREB3L1}.
\newblock PhD thesis, UT Southwestern Medical Center, 2013.

\bibitem{weinstein2013cancer}
John~N Weinstein, Eric~A Collisson, Gordon~B Mills, Kenna R~Mills Shaw, Brad~A
  Ozenberger, Kyle Ellrott, Ilya Shmulevich, Chris Sander, Joshua~M Stuart,
  Cancer Genome Atlas~Research Network, et~al.
\newblock The cancer genome atlas pan-cancer analysis project.
\newblock {\em Nature genetics}, 45(10):1113, 2013.

\end{thebibliography}

% \begin{appendix}
\section*{Appendix}
\subsection*{Proof of Theorem 1}
\begin{proof}
We first show that the sequence
$\mathcal{L}^f_{X,Y} (\boldsymbol{\theta}^{(m)}, \mathcal{C}^{(m)})$ is non-decreasing. Since $\boldsymbol{\theta}_k^{(m+1)}$ is maximizing $\prod_{i \in C_k^{(m)}} p(y_i, z_i = k| \boldsymbol{\theta}, \boldsymbol{x_i})$, then
% \begin{linenomath*}
\begin{equation*}
\mathcal{L}_{X,Y}^f (\boldsymbol{\theta}^{(m+1)}, \mathcal{C}^{(m)}) \geq \mathcal{L}^f_{X,Y} (\boldsymbol{\theta}^{(m)}, \mathcal{C}^{(m)})
\end{equation*}
% \end{linenomath*}
And since
% \begin{linenomath*}
\begin{equation*}
i \in \mathcal{C}_k^{(m+1)} \Leftrightarrow 
\mathcal{L}_{(\boldsymbol{x}_i, y_i)} (\boldsymbol{\theta}_k^{(m+1)}) 
\geq \mathcal{L}_{(\boldsymbol{x}_i,y_i)} (\boldsymbol{\theta}_{k'}^{(m+1)})
\end{equation*}
% \end{linenomath*}
for all $k' \neq k$, which implies
% \begin{linenomath*}
\begin{equation*}
\pi_k^{(m+1)} \mathcal{N} (y_i; \boldsymbol{x}_i^T \boldsymbol{\beta}_k^{(m+1)} , \sigma_k^{2^{(m+1)}}) \geq \pi_{k'}^{(m+1)} \mathcal{N} (y_i; \boldsymbol{x}_i^T \boldsymbol{\beta}_{k'}^{(m+1)} , \sigma_{k'}^{2^{(m+1)}})
\end{equation*}
% \end{linenomath*}
then we have
% \begin{linenomath*}
\begin{equation*}
\mathcal{L}^f_{X,Y} (\boldsymbol{\theta}^{(m+1)}, \mathcal{C}^{(m+1)})
\geq \mathcal{L}^f_{X,Y} (\boldsymbol{\theta}^{(m+1)}, \mathcal{C}^{(m)}) \geq \mathcal{L}^f_{X,Y} (\boldsymbol{\theta}^{(m)}, \mathcal{C}^{(m)})
\end{equation*}
% \end{linenomath*}
Since there is a finite number of partitions of the samples into $K$ clusters, the non-decreasing sequence $\mathcal{L}^f_{X,Y} (\boldsymbol{\theta}^{(m+1)}, \mathcal{C}^{m})$ takes a finite number of values, and thus, converges to a stationary values. Hence $\mathcal{L}^f_{X,Y} (\boldsymbol{\theta}^{(m)}, \mathcal{C}^{(m)}) 
= \mathcal{L}^f_{X,Y} (\boldsymbol{\theta}^{(m+1)}, \mathcal{C}^{(m)})
= \mathcal{L}^f_{X,Y} (\boldsymbol{\theta}^{(m+1)}, \mathcal{C}^{(m+1)})$ for $m$ large enough; from the first equality and from the assumption that the maximum likelihood estimate $\boldsymbol{\theta^{(m)}}$ are well-defined, we deduce that $\boldsymbol{\theta}^{(m)} = \boldsymbol{\theta}^{(m+1)}, \mathcal{C}^{(m)} = \mathcal{C}^{(m+1)}$.
\end{proof}

\subsection*{Proof of Theorem 2}
\begin{proof}
The proof is very similar to that of Theorem 1. We first show that the sequence $\mathcal{L}^{f,trim}_{X,Y} (\boldsymbol{\theta}^{(m)}, \mathcal{C}^{(m)})$ is non-decreasing. Since $\boldsymbol{\theta}_k^{(m+1)}$ is maximizing $\sum_{i=1}^{[n_k/2]+1}\{ (l_{k})_{i:n} + \log \pi_k\}$, then
% \begin{linenomath*}
\begin{equation*}
\mathcal{L}_{X,Y}^{f,trim} (\boldsymbol{\theta}^{(m+1)}, \mathcal{C}^{(m)}) \geq \mathcal{L}^{f,trim}_{X,Y} (\boldsymbol{\theta}^{(m)}, \mathcal{C}^{(m)})
\end{equation*}
% \end{linenomath*}
And since
% \begin{linenomath*}
\begin{equation*}
i \in \mathcal{C}_k^{(m+1)} \Leftrightarrow 
\mathcal{L}_{(\boldsymbol{x}_i, y_i)} (\boldsymbol{\theta}_k^{(m+1)}) 
\geq \mathcal{L}_{(\boldsymbol{x}_i,y_i)} (\boldsymbol{\theta}_{k'}^{(m+1)})
\end{equation*}
% \end{linenomath*}
for all $k' \neq k$, which implies
% \begin{linenomath*}
\begin{equation*}
\pi_k^{(m+1)} \mathcal{N} (y_i; \boldsymbol{x}_i^T \boldsymbol{\beta}_k^{(m+1)} , \sigma_k^{2^{(m+1)}}) \geq \pi_{k'}^{(m+1)} \mathcal{N} (y_i; \boldsymbol{x}_i^T \boldsymbol{\beta}_{k'}^{(m+1)} , \sigma_{k'}^{2^{(m+1)}})
\end{equation*}
% \end{linenomath*}
then we have
% \begin{linenomath*}
\begin{equation*}
\mathcal{L}^{f,trim}_{X,Y} (\boldsymbol{\theta}^{(m+1)}, \mathcal{C}^{(m+1)})
\geq \mathcal{L}^{f,trim}_{X,Y} (\boldsymbol{\theta}^{(m+1)}, \mathcal{C}^{(m)}) \geq \mathcal{L}^{f,trim}_{X,Y} (\boldsymbol{\theta}^{(m)}, \mathcal{C}^{(m)})
\end{equation*}
% \end{linenomath*}
Since there is a finite number of partitions of the samples into $K$ clusters, the non-decreasing sequence $\mathcal{L}^{f,trim}_{X,Y} (\boldsymbol{\theta}^{(m+1)}, \mathcal{C}^{m})$ takes a finite number of values, and thus, converges to a stationary values. Hence $\mathcal{L}^{f,trim}_{X,Y} (\boldsymbol{\theta}^{(m)}, \mathcal{C}^{(m)}) 
= \mathcal{L}^{f,trim}_{X,Y} (\boldsymbol{\theta}^{(m+1)}, \mathcal{C}^{(m)})
= \mathcal{L}^{f,trim}_{X,Y} (\boldsymbol{\theta}^{(m+1)}, \mathcal{C}^{(m+1)})$ for $m$ large enough; from the first equality and from the assumption that the maximum likelihood estimate $\boldsymbol{\theta^{(m)}}$ are well-defined, we deduce that $\boldsymbol{\theta}^{(m)} = \boldsymbol{\theta}^{(m+1)}, \mathcal{C}^{(m)} = \mathcal{C}^{(m+1)}$.
\end{proof}

\subsection*{Supplementary Tables}

% Supplementary Table 1
\begin{sidewaystable}
% \begin{table}
\renewcommand\thetable{S1}
\captionsetup{font=normalsize}
\caption{$K$=2, $P$=2, $N$=200}
\centering
\small
\begin{tabular}{ccccccccc}
\hline
Scenario & CAT           & MLE                  & TLE                  & CWM1             & CWM2          & MIXBI         & MIXL               & MIXT           \\ \hline
\multicolumn{9}{c}{Scenario 1: $\epsilon \sim {N}(0,1)$}   \\ \hline
1        & 0.010(0.170)  & -0.000(0.160)        & -0.000(0.160)        & 0.050(0.410)     & 0.000(0.470)  & 0.000(0.170)  & -0.050(0.230)      & -0.000(0.170)  \\
         & 0.030(0.130)  & 0.020(0.120)         & 0.020(0.120)         & -0.000(0.220)    & 0.150(1.280)  & 0.020(0.120)  & 0.030(0.140)       & 0.010(0.120)   \\
         & 0.010(0.160)  & 0.010(0.130)         & 0.010(0.130)         & 0.460(0.780)     & 0.780(1.020)  & 0.010(0.140)  & 0.070(0.230)       & 0.020(0.140)   \\
         & -0.000(0.120) & 0.000(0.100)         & 0.000(0.100)         & -0.100(0.370)    & -0.060(1.110) & 0.010(0.100)  & -0.010(0.130)      & 0.010(0.100)   \\
         & 0.000(0.210)  & -0.000(0.170)        & -0.000(0.170)        & 0.010(0.200)     & -0.030(0.530) & 0.000(0.180)  & -0.030(0.230)      & 0.010(0.190)   \\
         & 0.000(0.120)  & 0.010(0.110)         & 0.010(0.110)         & -0.010(0.250)    & 0.030(0.900)  & 0.010(0.110)  & 0.010(0.140)       & 0.010(0.120)   \\
         & 0.000(0.050)  & 0.000(0.050)         & 0.000(0.050)         & 0.040(0.120)     & 0.050(0.150)  & 0.000(0.050)  & -0.000(0.050)      & 0.000(0.050)   \\
         & -0.000(0.050) & -0.000(0.050)        & -0.000(0.050)        & -0.040(0.120)    & -0.050(0.150) & -0.000(0.050) & 0.000(0.050)       & -0.000(0.050)  \\ \hline
\multicolumn{9}{c}{Scenario 2: $\epsilon \sim {t}_1$}    \\ \hline
2        & 0.000(0.560)  & -202.260(2012.020)   & -202.260(2012.020)   & -20.150(279.560) & 0.270(2.540)  & 0.010(0.330)  & -571.250(5687.290) & 1.360(15.460)  \\
         & 0.030(0.230)  & 21.480(202.920)      & 21.480(202.920)      & 11.350(116.180)  & -0.370(2.560) & 0.010(0.200)  & -0.080(23.720)     & -3.410(29.910) \\
         & 0.450(0.690)  & -1200.700(11913.540) & -1200.700(11913.540) & -49.390(178.770) & 1.220(2.440)  & 1.180(0.800)  & -250.420(2486.920) & -5.820(27.590) \\
         & -0.020(0.230) & 38.960(160.180)      & 38.960(160.180)      & 30.400(109.180)  & -0.380(1.410) & 0.030(0.220)  & 5.710(27.100)      & 11.100(30.450) \\
         & -0.010(0.620) & -152.190(1483.380)   & -152.190(1483.380)   & -20.880(297.780) & -0.170(3.470) & -0.000(0.330) & 196.580(1964.350)  & -1.490(17.740) \\
         & -0.010(0.230) & 32.330(156.300)      & 32.330(156.300)      & 19.390(231.390)  & -0.160(2.420) & -0.020(0.220) & -1.690(28.180)     & -0.850(25.270) \\
         & 0.110(0.110)  & 0.130(0.420)         & 0.130(0.420)         & 0.110(0.420)     & 0.080(0.200)  & 0.250(0.130)  & 0.040(0.350)       & 0.150(0.410)   \\
         & -0.110(0.110) & -0.130(0.420)        & -0.130(0.420)        & -0.110(0.420)    & -0.080(0.200) & -0.250(0.130) & -0.040(0.350)      & -0.150(0.410)  \\ \hline
\multicolumn{9}{c}{Scenario 3: $\epsilon \sim {t}_3$}    \\ \hline
3        & -0.040(0.170) & -0.020(0.260)        & -0.020(0.260)        & 0.020(1.060)     & -0.050(0.530) & -0.040(0.170) & -0.100(0.260)      & -0.010(0.270)  \\
         & -0.020(0.150) & 0.090(1.100)         & 0.090(1.100)         & -0.010(0.260)    & -0.060(0.430) & -0.020(0.150) & -0.030(0.160)      & -0.020(0.160)  \\
         & 0.050(0.180)  & 0.370(0.580)         & 0.370(0.580)         & 0.920(1.110)     & 0.800(0.980)  & 0.050(0.240)  & 0.080(0.220)       & 0.500(0.610)   \\
         & -0.010(0.140) & 0.220(1.990)         & 0.220(1.990)         & -0.250(0.550)    & -0.080(0.700) & 0.020(0.140)  & -0.020(0.170)      & 0.020(0.150)   \\
         & 0.030(0.220)  & 0.000(0.340)         & 0.000(0.340)         & -0.120(0.840)    & 0.040(0.380)  & 0.000(0.220)  & -0.050(0.260)      & 0.020(0.350)   \\
         & -0.010(0.140) & 0.150(1.510)         & 0.150(1.510)         & 0.020(0.320)     & -0.010(0.540) & -0.010(0.130) & -0.010(0.150)      & 0.010(0.160)   \\
         & 0.020(0.070)  & 0.050(0.090)         & 0.050(0.090)         & 0.060(0.160)     & 0.090(0.160)  & 0.020(0.080)  & 0.000(0.060)       & 0.070(0.080)   \\
         & -0.020(0.070) & -0.050(0.090)        & -0.050(0.090)        & -0.060(0.160)    & -0.090(0.160) & -0.020(0.080) & -0.000(0.060)      & -0.070(0.080)  \\ \hline
\multicolumn{9}{c}{Scenario 4: 5\% added outliers}    \\ \hline
4        & -0.070(0.190) & -0.180(0.520)        & -0.120(1.270)        & -0.030(1.620)    & 0.050(0.790)  & -0.170(0.200) & -0.270(0.330)      & 0.250(0.430)   \\
         & 0.020(0.100)  & 0.330(0.330)         & -0.030(1.310)        & -0.130(1.990)    & 0.320(4.680)  & 0.050(0.090)  & 0.120(0.130)       & 0.030(0.160)   \\
         & -0.120(0.210) & 0.230(0.890)         & 2.110(2.250)         & 0.620(1.170)     & 0.820(1.180)  & -0.250(0.190) & -0.250(0.330)      & 0.970(0.640)   \\
         & 0.040(0.120)  & 0.190(0.640)         & -0.930(2.170)        & -0.030(1.000)    & -0.040(0.610) & 0.080(0.120)  & 0.170(0.170)       & -0.240(0.420)  \\
         & -0.190(0.260) & -0.540(1.060)        & -0.450(2.090)        & 0.350(2.140)     & 0.070(1.210)  & -0.410(0.270) & -0.560(0.550)      & 0.540(0.860)   \\
         & 0.030(0.130)  & 0.690(0.660)         & 0.040(1.480)         & 0.200(0.920)     & -0.020(3.580) & 0.100(0.150)  & 0.260(0.190)       & 0.070(0.200)   \\
         & 0.020(0.050)  & 0.010(0.130)         & 0.090(0.160)         & 0.040(0.190)     & 0.090(0.220)  & 0.020(0.040)  & -0.000(0.050)      & -0.020(0.120)  \\
         & -0.020(0.050) & -0.010(0.130)        & -0.090(0.160)        & -0.040(0.190)    & -0.090(0.220) & -0.020(0.040) & 0.000(0.050)       & 0.020(0.120)   \\ \hline
\multicolumn{9}{c}{Scenario 5: 10\% added outliers}  \\ \hline
5        & -0.030(0.170) & 0.020(0.680)         & -0.120(1.470)        & -0.030(0.580)    & -0.010(0.990) & -0.050(0.160) & -0.120(0.230)      & 0.170(0.410)   \\
         & 0.010(0.110)  & 0.170(0.360)         & -0.230(1.530)        & -0.020(0.320)    & 0.040(0.200)  & 0.020(0.110)  & 0.070(0.150)       & 0.010(0.140)   \\
         & -0.020(0.150) & 0.520(0.840)         & 2.520(2.590)         & 0.250(0.700)     & 0.230(0.950)  & -0.060(0.150) & -0.060(0.200)      & 0.680(0.640)   \\
         & -0.010(0.120) & -0.010(0.490)        & -0.970(2.370)        & -0.110(0.460)    & -0.080(0.360) & 0.010(0.120)  & 0.030(0.140)       & -0.130(0.270)  \\
         & -0.060(0.190) & 0.030(1.170)         & -0.070(2.110)        & 0.010(0.760)     & 0.010(1.730)  & -0.140(0.200) & -0.270(0.230)      & 0.400(0.650)   \\
         & -0.010(0.110) & 0.340(0.670)         & 0.030(2.130)         & 0.020(0.170)     & 0.030(0.390)  & 0.010(0.100)  & 0.110(0.150)       & 0.010(0.200)   \\
         & -0.010(0.050) & -0.040(0.120)        & 0.120(0.080)         & 0.020(0.130)     & 0.010(0.150)  & -0.000(0.040) & -0.010(0.050)      & -0.040(0.080)  \\
         & 0.010(0.050)  & 0.040(0.120)         & -0.120(0.080)        & -0.020(0.130)    & -0.010(0.150) & 0.000(0.040)  & 0.010(0.050)       & 0.040(0.080)   \\ \hline
\end{tabular}
% \end{table}
\end{sidewaystable}

% Supplementary Table 2
\begin{sidewaystable}
% \begin{table}
\renewcommand\thetable{S2}
\captionsetup{font=normalsize}
\caption{$K$=2, $P$=2, $N$=400}
\centering
\small
\begin{tabular}{ccccccccc}
\hline
Scenario & CAT           & MLE             & TLE             & CWM1             & CWM2          & MIXBI         & MIXL           & MIXT           \\ \hline
\multicolumn{9}{c}{Scenario 1: $\epsilon \sim {N}(0,1)$}   \\ \hline
1        & -0.010(0.120) & -0.010(0.100)   & -0.010(0.100)   & 0.010(0.110)     & -0.040(0.190) & -0.010(0.110) & -0.020(0.140)  & -0.010(0.100)  \\
         & 0.000(0.080)  & 0.010(0.070)    & 0.010(0.070)    & -0.000(0.090)    & 0.010(0.130)  & 0.000(0.080)  & 0.000(0.090)   & 0.010(0.070)   \\
         & 0.010(0.110)  & 0.010(0.100)    & 0.010(0.100)    & 0.310(0.660)     & 0.580(0.880)  & 0.010(0.100)  & 0.010(0.120)   & 0.010(0.110)   \\
         & 0.000(0.070)  & 0.010(0.060)    & 0.010(0.060)    & -0.080(0.300)    & -0.030(0.260) & 0.010(0.060)  & -0.010(0.080)  & 0.020(0.070)   \\
         & -0.010(0.110) & -0.010(0.100)   & -0.010(0.100)   & -0.010(0.210)    & 0.020(0.160)  & -0.010(0.100) & 0.000(0.110)   & -0.010(0.110)  \\
         & 0.000(0.070)  & 0.010(0.070)    & 0.010(0.070)    & -0.010(0.110)    & -0.020(0.130) & 0.010(0.070)  & 0.000(0.080)   & 0.010(0.080)   \\
         & 0.010(0.040)  & 0.010(0.030)    & 0.010(0.030)    & 0.020(0.090)     & 0.070(0.140)  & 0.010(0.030)  & 0.000(0.030)   & 0.010(0.030)   \\
         & -0.010(0.040) & -0.010(0.030)   & -0.010(0.030)   & -0.020(0.090)    & -0.070(0.140) & -0.010(0.030) & -0.000(0.030)  & -0.010(0.030)  \\ \hline
\multicolumn{9}{c}{Scenario 2: $\epsilon \sim {t}_1$}    \\ \hline
2        & 0.040(0.430)  & 4.000(58.690)   & 4.000(58.690)   & -3.580(44.630)   & -0.100(1.020) & -0.000(0.270) & -0.170(14.690) & -0.450(12.130) \\
         & -0.010(0.110) & -11.560(89.090) & -11.560(89.090) & -20.980(265.540) & 0.110(0.580)  & -0.000(0.120) & 9.410(77.350)  & -2.160(19.060) \\
         & 0.440(0.500)  & -8.230(34.550)  & -8.230(34.550)  & -14.000(40.110)  & 1.430(1.620)  & 1.240(0.590)  & -2.780(14.890) & -2.570(8.740)  \\
         & -0.020(0.160) & 18.410(65.280)  & 18.410(65.280)  & 37.620(213.420)  & -0.540(0.580) & 0.030(0.140)  & 16.120(84.360) & 3.510(9.710)   \\
         & 0.020(0.280)  & -5.230(62.060)  & -5.230(62.060)  & -1.150(98.320)   & -0.230(1.780) & -0.010(0.220) & -0.930(9.590)  & -2.280(14.630) \\
         & -0.000(0.100) & -5.450(77.360)  & -5.450(77.360)  & -11.880(60.880)  & -0.050(1.150) & -0.000(0.120) & -0.650(57.100) & 0.310(13.440)  \\
         & 0.120(0.090)  & 0.160(0.430)    & 0.160(0.430)    & 0.080(0.440)     & 0.100(0.210)  & 0.270(0.100)  & 0.080(0.390)   & 0.090(0.420)   \\
         & -0.120(0.090) & -0.160(0.430)   & -0.160(0.430)   & -0.080(0.440)    & -0.100(0.210) & -0.270(0.100) & -0.080(0.390)  & -0.090(0.420)  \\ \hline
\multicolumn{9}{c}{Scenario 3: $\epsilon \sim {t}_3$}    \\ \hline
3        & -0.000(0.130) & 0.020(0.190)    & 0.020(0.190)    & -0.030(0.220)    & -0.030(0.320) & 0.010(0.120)  & -0.010(0.140)  & 0.010(0.200)   \\
         & 0.000(0.100)  & -0.000(0.110)   & -0.000(0.110)   & 0.010(0.170)     & 0.000(0.320)  & -0.000(0.090) & 0.010(0.100)   & -0.000(0.100)  \\
         & 0.030(0.120)  & 0.430(0.440)    & 0.430(0.440)    & 0.970(0.800)     & 0.830(0.980)  & 0.020(0.120)  & 0.030(0.120)   & 0.530(0.450)   \\
         & -0.000(0.090) & 0.040(0.100)    & 0.040(0.100)    & -0.150(0.410)    & -0.140(0.360) & 0.020(0.090)  & -0.010(0.090)  & 0.040(0.110)   \\
         & -0.020(0.130) & 0.010(0.190)    & 0.010(0.190)    & 0.020(0.230)     & 0.020(0.280)  & -0.020(0.130) & -0.030(0.140)  & 0.000(0.220)   \\
         & 0.010(0.100)  & 0.010(0.100)    & 0.010(0.100)    & 0.010(0.390)     & 0.000(0.200)  & 0.010(0.090)  & 0.010(0.100)   & 0.010(0.100)   \\
         & 0.020(0.040)  & 0.040(0.050)    & 0.040(0.050)    & 0.080(0.140)     & 0.070(0.160)  & 0.020(0.040)  & -0.000(0.040)  & 0.060(0.060)   \\
         & -0.020(0.040) & -0.040(0.050)   & -0.040(0.050)   & -0.080(0.140)    & -0.070(0.160) & -0.020(0.040) & 0.000(0.040)   & -0.060(0.060)  \\ \hline
\multicolumn{9}{c}{Scenario 4: 5\% added outliers}    \\ \hline
4        & -0.120(0.130) & -0.140(0.530)   & 0.040(0.790)    & -0.330(3.180)    & -0.050(0.550) & -0.230(0.140) & -0.310(0.170)  & 0.320(0.270)   \\
         & 0.030(0.090)  & 0.280(0.280)    & 0.050(0.690)    & -0.130(1.760)    & -0.560(4.150) & 0.070(0.100)  & 0.160(0.130)   & 0.030(0.100)   \\
         & -0.150(0.160) & 0.300(0.750)    & 2.440(2.180)    & 0.400(0.990)     & 0.770(1.240)  & -0.290(0.140) & -0.330(0.190)  & 1.040(0.360)   \\
         & 0.010(0.090)  & 0.230(0.400)    & -0.960(1.650)   & 0.070(0.990)     & 0.270(2.060)  & 0.060(0.100)  & 0.140(0.140)   & -0.190(0.200)  \\
         & -0.190(0.210) & -0.280(1.020)   & -0.040(0.860)   & 0.220(2.490)     & 0.010(0.570)  & -0.410(0.180) & -0.610(0.270)  & 0.730(0.450)   \\
         & 0.030(0.090)  & 0.590(0.580)    & 0.090(0.820)    & 0.220(0.910)     & 0.080(1.810)  & 0.110(0.100)  & 0.280(0.160)   & 0.050(0.100)   \\
         & 0.010(0.030)  & -0.000(0.060)   & 0.100(0.170)    & 0.040(0.170)     & 0.060(0.200)  & 0.000(0.030)  & -0.010(0.040)  & -0.030(0.050)  \\
         & -0.010(0.030) & 0.000(0.060)    & -0.100(0.170)   & -0.040(0.170)    & -0.060(0.200) & -0.000(0.030) & 0.010(0.040)   & 0.030(0.050)   \\ \hline
\multicolumn{9}{c}{Scenario 5: 10\% added outliers}  \\ \hline
5        & -0.060(0.110) & 0.030(0.410)    & -0.210(1.510)   & -0.010(0.180)    & 0.030(0.470)  & -0.090(0.120) & -0.140(0.130)  & 0.180(0.280)   \\
         & 0.020(0.070)  & 0.150(0.270)    & -0.090(1.350)   & -0.070(0.740)    & 0.010(0.150)  & 0.030(0.070)  & 0.080(0.100)   & 0.020(0.090)   \\
         & -0.060(0.110) & 0.520(0.640)    & 2.570(2.000)    & 0.310(0.800)     & 0.500(0.840)  & -0.110(0.120) & -0.120(0.140)  & 0.660(0.520)   \\
         & -0.000(0.080) & -0.000(0.330)   & -0.910(2.110)   & -0.060(0.240)    & -0.070(0.320) & 0.020(0.080)  & 0.050(0.100)   & -0.130(0.160)  \\
         & -0.080(0.150) & 0.010(0.720)    & 0.220(1.800)    & -0.040(0.220)    & -0.050(0.310) & -0.160(0.160) & -0.260(0.170)  & 0.390(0.540)   \\
         & 0.020(0.090)  & 0.290(0.410)    & 0.320(1.720)    & -0.050(0.680)    & -0.010(0.150) & 0.050(0.090)  & 0.130(0.100)   & 0.030(0.110)   \\
         & 0.000(0.040)  & -0.020(0.100)   & 0.140(0.160)    & 0.030(0.110)     & 0.040(0.130)  & 0.000(0.030)  & -0.000(0.030)  & -0.040(0.060)  \\
         & -0.000(0.040) & 0.020(0.100)    & -0.140(0.160)   & -0.030(0.110)    & -0.040(0.130) & -0.000(0.030) & 0.000(0.030)   & 0.040(0.060)   \\ \hline
\end{tabular}
% \end{table}
\end{sidewaystable}

% Supplementary Table 3
\begin{sidewaystable}
% \begin{table}
\renewcommand\thetable{S3}
\captionsetup{font=normalsize}
\caption{$K$=3, $P$=1, $N$=200}
\centering
\small
\begin{tabular}{ccccccccc}
\hline
Scenario & CAT           & MLE              & TLE              & CWM1             & CWM2          & MIXBI         & MIXL             & MIXT           \\ \hline
\multicolumn{9}{c}{Scenario 1: $\epsilon \sim {N}(0,1)$}   \\ \hline
1        & -0.220(0.620) & -0.110(0.550)    & -0.110(0.550)    & -0.750(0.600)    & -0.800(0.640) & -0.150(0.510) & -0.440(0.500)    & -0.260(0.580)  \\
         & 0.020(0.220)  & -0.010(0.190)    & -0.010(0.190)    & 0.010(0.420)     & -0.090(0.530) & -0.000(0.180) & 0.030(0.220)     & 0.010(0.220)   \\
         & -0.070(0.350) & 0.070(0.360)     & 0.070(0.360)     & 0.250(0.660)     & 0.130(0.490)  & 0.030(0.280)  & 0.060(0.270)     & 0.100(0.430)   \\
         & 0.290(0.470)  & 0.090(0.580)     & 0.090(0.580)     & 0.590(0.890)     & 0.810(1.150)  & 0.050(0.340)  & 0.060(0.410)     & 0.100(0.600)   \\
         & -0.030(0.250) & 0.010(0.210)     & 0.010(0.210)     & -0.240(0.740)    & -0.430(0.810) & 0.010(0.190)  & -0.040(0.210)    & 0.020(0.300)   \\
         & 0.060(0.380)  & 0.130(0.530)     & 0.130(0.530)     & 0.280(0.910)     & 0.170(0.620)  & 0.080(0.370)  & 0.090(0.240)     & 0.260(0.720)   \\
         & 0.050(0.110)  & 0.030(0.120)     & 0.030(0.120)     & 0.120(0.160)     & 0.120(0.130)  & 0.040(0.130)  & 0.050(0.100)     & 0.060(0.140)   \\
         & -0.000(0.060) & -0.000(0.060)    & -0.000(0.060)    & -0.000(0.110)    & 0.010(0.120)  & -0.000(0.060) & 0.000(0.050)     & -0.010(0.070)  \\
         & -0.050(0.110) & -0.030(0.100)    & -0.030(0.100)    & -0.120(0.150)    & -0.140(0.130) & -0.030(0.120) & -0.050(0.100)    & -0.060(0.120)  \\ \hline
\multicolumn{9}{c}{Scenario 2: $\epsilon \sim {t}_1$}    \\ \hline
2        & -0.160(1.280) & 8.210(39.220)    & 8.210(39.220)    & 13.380(89.170)   & 0.050(4.260)  & -0.550(0.760) & 34.580(192.760)  & 0.630(7.860)   \\
         & -0.070(1.190) & 48.250(303.090)  & 48.250(303.090)  & -3.880(238.410)  & -0.630(2.030) & -0.310(0.620) & 33.300(252.260)  & 0.720(4.580)   \\
         & 0.170(0.840)  & -34.590(162.640) & -34.590(162.640) & -18.070(47.120)  & -0.100(2.170) & 0.090(0.500)  & -21.280(82.890)  & -4.690(13.370) \\
         & 0.660(1.840)  & -45.270(336.280) & -45.270(336.280) & -36.040(141.380) & -0.310(8.410) & 0.820(0.990)  & -55.330(362.950) & -3.550(13.840) \\
         & 0.330(1.480)  & 73.070(389.780)  & 73.070(389.780)  & 70.780(326.440)  & -0.500(1.670) & -0.450(0.870) & 34.570(179.730)  & 2.160(5.750)   \\
         & 0.330(1.020)  & -20.300(122.470) & -20.300(122.470) & -4.680(35.450)   & 0.250(1.880)  & 0.190(0.580)  & -8.940(48.910)   & -4.050(15.890) \\
         & 0.160(0.160)  & 0.160(0.250)     & 0.160(0.250)     & 0.140(0.280)     & 0.110(0.160)  & 0.200(0.210)  & 0.140(0.300)     & 0.170(0.230)   \\
         & -0.070(0.080) & -0.050(0.210)    & -0.050(0.210)    & 0.000(0.270)     & 0.020(0.160)  & 0.000(0.210)  & -0.050(0.220)    & -0.020(0.200)  \\
         & -0.090(0.150) & -0.110(0.270)    & -0.110(0.270)    & -0.140(0.260)    & -0.130(0.160) & -0.210(0.140) & -0.090(0.310)    & -0.150(0.220)  \\ \hline
\multicolumn{9}{c}{Scenario 3: $\epsilon \sim {t}_3$}    \\ \hline
3        & -0.060(0.850) & -0.630(1.160)    & -0.630(1.160)    & -0.770(0.760)    & -0.680(0.700) & -0.680(0.630) & -0.520(0.580)    & -0.660(0.620)  \\
         & 0.010(0.300)  & 0.100(2.600)     & 0.100(2.600)     & -0.120(0.570)    & -0.150(0.520) & -0.030(0.290) & 0.010(0.430)     & -0.040(0.450)  \\
         & -0.090(0.370) & -0.340(2.080)    & -0.340(2.080)    & 0.190(0.700)     & 0.090(0.400)  & -0.030(0.330) & 0.000(0.690)     & 0.150(0.530)   \\
         & 0.290(0.670)  & 0.090(1.760)     & 0.090(1.760)     & 0.610(1.170)     & 0.880(0.980)  & 0.490(0.750)  & 0.060(0.740)     & 0.510(0.820)   \\
         & -0.040(0.290) & 0.080(1.190)     & 0.080(1.190)     & -0.370(0.770)    & -0.560(0.840) & 0.050(0.300)  & -0.030(0.280)    & -0.050(0.520)  \\
         & 0.080(0.300)  & 0.300(1.580)     & 0.300(1.580)     & 0.280(0.960)     & 0.250(0.700)  & 0.220(0.470)  & 0.130(0.800)     & 0.400(0.850)   \\
         & 0.060(0.100)  & 0.130(0.170)     & 0.130(0.170)     & 0.130(0.150)     & 0.140(0.130)  & 0.210(0.170)  & 0.070(0.130)     & 0.150(0.150)   \\
         & -0.010(0.070) & -0.010(0.090)    & -0.010(0.090)    & 0.000(0.120)     & 0.020(0.140)  & -0.030(0.090) & -0.000(0.070)    & -0.020(0.100)  \\
         & -0.050(0.110) & -0.120(0.160)    & -0.120(0.160)    & -0.130(0.150)    & -0.150(0.140) & -0.170(0.140) & -0.070(0.120)    & -0.140(0.130)  \\ \hline
\multicolumn{9}{c}{Scenario 4: 5\% added outliers}    \\ \hline
4        & -0.470(0.770) & -1.510(1.370)    & -0.870(1.240)    & 0.560(9.140)     & -0.620(0.760) & -1.580(0.990) & -2.280(1.110)    & -0.670(1.090)  \\
         & 0.140(0.410)  & 0.560(0.580)     & -0.520(1.360)    & -0.300(1.120)    & -0.150(0.510) & 0.300(0.500)  & 0.320(0.350)     & 0.200(0.330)   \\
         & 0.110(0.300)  & -0.000(1.350)    & 0.880(1.200)     & 0.470(0.620)     & 0.070(1.490)  & 0.160(0.580)  & 0.090(0.940)     & 0.400(0.580)   \\
         & -0.290(1.160) & -0.750(2.360)    & 1.170(2.520)     & -0.420(5.370)    & 0.850(1.160)  & -0.770(1.760) & -2.080(2.090)    & 0.620(1.530)   \\
         & 0.220(0.470)  & 0.750(0.780)     & -0.960(2.240)    & -0.050(1.270)    & -0.330(0.960) & 0.510(0.550)  & 0.520(0.490)     & 0.260(0.580)   \\
         & 0.120(0.420)  & -0.640(1.650)    & 0.810(2.060)     & 0.420(1.060)     & 0.280(1.330)  & 0.130(0.950)  & -0.300(1.300)    & -0.300(0.790)  \\
         & 0.020(0.090)  & 0.020(0.220)     & 0.080(0.140)     & 0.110(0.150)     & 0.090(0.150)  & 0.140(0.170)  & -0.020(0.160)    & -0.030(0.130)  \\
         & -0.010(0.070) & -0.040(0.080)    & 0.010(0.150)     & -0.010(0.140)    & 0.010(0.140)  & 0.010(0.080)  & 0.020(0.080)     & -0.030(0.060)  \\
         & -0.010(0.100) & 0.010(0.200)     & -0.090(0.160)    & -0.090(0.140)    & -0.100(0.170) & -0.140(0.150) & -0.000(0.160)    & 0.060(0.140)   \\ \hline
\multicolumn{9}{c}{Scenario 5: 10\% added outliers}  \\ \hline
5        & -0.290(0.610) & -0.930(1.530)    & -0.790(1.410)    & -0.490(1.400)    & -0.720(0.710) & -0.930(0.720) & -1.440(1.130)    & -0.650(1.310)  \\
         & 0.050(0.200)  & 0.350(0.490)     & -0.870(1.980)    & -0.210(0.730)    & -0.060(0.490) & 0.070(0.220)  & 0.340(0.750)     & 0.160(0.410)   \\
         & -0.040(0.280) & 0.080(1.400)     & 1.080(1.380)     & 0.300(0.640)     & 0.230(0.570)  & 0.060(0.260)  & 0.210(0.880)     & 0.500(0.620)   \\
         & 0.120(0.590)  & 0.050(2.280)     & 1.270(1.920)     & 0.630(1.520)     & 0.740(1.050)  & -0.100(1.250) & -0.770(2.080)    & 0.500(1.790)   \\
         & 0.030(0.210)  & 0.340(0.470)     & -1.390(2.630)    & -0.180(1.050)    & -0.430(0.610) & 0.120(0.250)  & 0.330(0.590)     & 0.150(0.430)   \\
         & 0.050(0.170)  & -0.520(1.700)    & 0.520(1.910)     & 0.200(0.950)     & 0.300(0.730)  & 0.150(0.350)  & -0.020(1.400)    & -0.090(1.050)  \\
         & 0.040(0.100)  & 0.020(0.250)     & 0.110(0.190)     & 0.120(0.150)     & 0.130(0.150)  & 0.130(0.140)  & 0.030(0.190)     & -0.050(0.180)  \\
         & 0.010(0.050)  & -0.030(0.100)    & 0.030(0.140)     & -0.010(0.110)    & 0.000(0.130)  & 0.020(0.060)  & -0.010(0.110)    & -0.020(0.080)  \\
         & -0.050(0.110) & 0.010(0.240)     & -0.140(0.200)    & -0.110(0.140)    & -0.140(0.150) & -0.150(0.130) & -0.030(0.160)    & 0.070(0.160)   \\ \hline
\end{tabular}
% \end{table}
\end{sidewaystable}

% Supplementary Table 4
\begin{sidewaystable}
% \begin{table}
\renewcommand\thetable{S4}
\captionsetup{font=normalsize}
\caption{$K$=3, $P$=1, $N$=400}
\centering
\small
\begin{tabular}{ccccccccc}
\hline
Scenario & CAT           & MLE                & TLE                & CWM1              & CWM2          & MIXBI         & MIXL             & MIXT           \\ \hline
\multicolumn{9}{c}{Scenario 1: $\epsilon \sim {N}(0,1)$}   \\ \hline
1        & -0.030(0.420) & -0.000(0.360)      & -0.000(0.360)      & -0.650(0.590)     & -0.710(0.690) & 0.030(0.280)  & -0.220(0.390)    & 0.010(0.290)   \\
         & 0.030(0.130)  & 0.020(0.160)       & 0.020(0.160)       & 0.010(0.260)      & -0.050(0.410) & 0.010(0.110)  & 0.030(0.150)     & 0.020(0.120)   \\
         & -0.050(0.210) & 0.040(0.330)       & 0.040(0.330)       & 0.310(0.540)      & 0.110(0.410)  & -0.010(0.140) & 0.020(0.190)     & 0.010(0.150)   \\
         & 0.220(0.350)  & 0.000(0.220)       & 0.000(0.220)       & 0.610(0.740)      & 1.050(0.910)  & -0.000(0.200) & -0.000(0.290)    & -0.080(0.220)  \\
         & 0.010(0.130)  & 0.020(0.130)       & 0.020(0.130)       & -0.200(0.430)     & -0.360(0.560) & 0.020(0.100)  & -0.010(0.140)    & 0.010(0.120)   \\
         & -0.010(0.130) & 0.080(0.480)       & 0.080(0.480)       & 0.350(0.750)      & 0.170(0.450)  & -0.010(0.110) & 0.030(0.130)     & 0.020(0.120)   \\
         & 0.010(0.080)  & 0.010(0.090)       & 0.010(0.090)       & 0.130(0.150)      & 0.120(0.150)  & 0.000(0.050)  & 0.020(0.070)     & -0.010(0.050)  \\
         & -0.000(0.040) & -0.010(0.050)      & -0.010(0.050)      & -0.010(0.090)     & -0.000(0.110) & -0.000(0.040) & -0.000(0.040)    & -0.000(0.030)  \\
         & -0.000(0.090) & -0.000(0.070)      & -0.000(0.070)      & -0.130(0.130)     & -0.120(0.130) & 0.000(0.060)  & -0.020(0.070)    & 0.010(0.060)   \\ \hline
\multicolumn{9}{c}{Scenario 2: $\epsilon \sim {t}_1$}    \\ \hline
2        & -0.530(1.130) & -4.720(240.250)    & -4.720(240.250)    & 12.570(59.800)    & -0.420(2.120) & -0.670(0.510) & 29.950(107.900)  & 1.520(7.220)   \\
         & 0.030(0.280)  & -3.670(124.420)    & -3.670(124.420)    & 123.370(1149.950) & -0.450(2.220) & -0.260(0.400) & 16.420(219.360)  & 0.320(3.850)   \\
         & 0.120(0.800)  & -52.110(299.120)   & -52.110(299.120)   & -55.660(461.180)  & -0.010(1.560) & 0.080(0.230)  & -68.210(227.540) & -4.220(12.490) \\
         & 0.640(1.290)  & -169.330(1179.540) & -169.330(1179.540) & -21.540(103.290)  & 0.680(2.290)  & 0.820(0.830)  & -47.840(216.920) & -1.350(7.120)  \\
         & 0.000(0.290)  & 28.510(104.000)    & 28.510(104.000)    & 81.240(445.560)   & -0.590(1.820) & -0.560(0.800) & 77.830(353.090)  & 1.240(4.470)   \\
         & 0.280(0.800)  & -23.250(235.490)   & -23.250(235.490)   & -21.000(214.110)  & 0.180(1.120)  & 0.060(0.220)  & -31.740(134.970) & -2.310(9.870)  \\
         & 0.170(0.170)  & 0.130(0.290)       & 0.130(0.290)       & 0.130(0.270)      & 0.120(0.170)  & 0.190(0.230)  & 0.150(0.360)     & 0.180(0.230)   \\
         & -0.060(0.090) & -0.000(0.270)      & -0.000(0.270)      & -0.060(0.250)     & 0.010(0.160)  & 0.040(0.240)  & -0.070(0.290)    & -0.030(0.190)  \\
         & -0.110(0.160) & -0.120(0.310)      & -0.120(0.310)      & -0.070(0.280)     & -0.130(0.170) & -0.230(0.110) & -0.090(0.360)    & -0.150(0.220)  \\ \hline
\multicolumn{9}{c}{Scenario 3: $\epsilon \sim {t}_3$}    \\ \hline
3        & 0.020(0.470)  & -0.340(2.120)      & -0.340(2.120)      & -0.670(0.780)     & -0.800(0.580) & -0.740(0.540) & -0.330(0.410)    & -0.620(0.610)  \\
         & 0.020(0.250)  & -0.050(1.360)      & -0.050(1.360)      & -0.250(0.600)     & -0.100(0.440) & -0.000(0.200) & 0.010(0.190)     & -0.040(0.290)  \\
         & -0.060(0.240) & 0.260(0.580)       & 0.260(0.580)       & 0.360(0.560)      & 0.250(0.500)  & -0.010(0.310) & -0.010(0.320)    & 0.280(0.450)   \\
         & 0.290(0.360)  & 0.490(0.760)       & 0.490(0.760)       & 1.030(0.940)      & 0.790(0.830)  & 0.440(0.490)  & -0.000(0.360)    & 0.750(0.780)   \\
         & -0.020(0.190) & 0.060(1.400)       & 0.060(1.400)       & -0.540(0.680)     & -0.460(0.620) & 0.030(0.150)  & -0.040(0.170)    & -0.070(0.300)  \\
         & 0.040(0.370)  & 0.230(0.820)       & 0.230(0.820)       & 0.270(0.840)      & 0.250(0.610)  & 0.150(0.480)  & 0.100(0.380)     & 0.210(0.660)   \\
         & 0.050(0.090)  & 0.100(0.170)       & 0.100(0.170)       & 0.140(0.150)      & 0.120(0.140)  & 0.230(0.170)  & 0.040(0.080)     & 0.120(0.150)   \\
         & -0.010(0.050) & -0.010(0.090)      & -0.010(0.090)      & -0.010(0.110)     & 0.010(0.120)  & -0.030(0.060) & 0.010(0.050)     & -0.030(0.070)  \\
         & -0.040(0.090) & -0.090(0.140)      & -0.090(0.140)      & -0.130(0.140)     & -0.140(0.130) & -0.200(0.130) & -0.040(0.090)    & -0.090(0.150)  \\ \hline
\multicolumn{9}{c}{Scenario 4: 5\% added outliers}    \\ \hline
4        & -0.720(0.740) & -1.700(1.340)      & -1.060(0.960)      & 0.340(6.670)      & -0.550(0.930) & -1.420(0.900) & -2.520(1.150)    & -0.630(0.770)  \\
         & 0.090(0.200)  & 0.530(0.390)       & -0.760(1.100)      & -0.470(1.060)     & -0.120(0.430) & 0.250(0.200)  & 0.380(0.340)     & 0.170(0.260)   \\
         & 0.090(0.210)  & 0.080(1.080)       & 0.600(1.040)       & 0.510(0.640)      & 0.210(0.780)  & 0.090(0.320)  & 0.030(1.040)     & 0.520(0.210)   \\
         & -0.500(0.750) & -0.810(2.120)      & 1.600(1.830)       & -0.150(4.160)     & 0.980(1.160)  & -0.600(1.260) & -2.210(2.100)    & 0.650(1.190)   \\
         & 0.110(0.240)  & 0.690(0.590)       & -1.340(2.040)      & 0.210(1.300)      & -0.360(0.680) & 0.410(0.240)  & 0.450(0.370)     & 0.240(0.450)   \\
         & 0.070(0.150)  & -0.830(1.600)      & 0.720(1.820)       & 0.380(0.980)      & 0.020(0.770)  & 0.070(0.540)  & -0.470(1.680)    & -0.330(0.250)  \\
         & 0.040(0.070)  & -0.010(0.190)      & 0.100(0.170)       & 0.140(0.170)      & 0.080(0.150)  & 0.130(0.120)  & 0.000(0.190)     & -0.050(0.120)  \\
         & 0.010(0.040)  & -0.030(0.050)      & 0.010(0.180)       & -0.040(0.120)     & 0.010(0.130)  & 0.030(0.040)  & 0.020(0.070)     & -0.030(0.040)  \\
         & -0.050(0.080) & 0.040(0.200)       & -0.110(0.160)      & -0.100(0.140)     & -0.090(0.150) & -0.160(0.110) & -0.020(0.170)    & 0.080(0.120)   \\ \hline
\multicolumn{9}{c}{Scenario 5: 10\% added outliers}  \\ \hline
5        & -0.200(0.500) & -1.260(1.680)      & -1.040(1.320)      & -0.280(2.780)     & -0.720(0.580) & -0.870(0.290) & -1.420(1.050)    & -0.390(1.200)  \\
         & 0.010(0.130)  & 0.250(0.310)       & -0.700(1.860)      & -0.200(0.570)     & -0.130(0.530) & 0.040(0.120)  & 0.260(0.440)     & 0.120(0.280)   \\
         & 0.020(0.250)  & 0.360(0.830)       & 0.320(1.360)       & 0.300(1.400)      & 0.290(0.540)  & 0.030(0.150)  & 0.330(0.730)     & 0.530(0.380)   \\
         & 0.040(0.440)  & -0.480(2.180)      & 0.900(2.490)       & 0.380(2.170)      & 1.010(1.070)  & 0.100(0.390)  & -0.700(1.880)    & 0.730(1.520)   \\
         & 0.030(0.160)  & 0.290(0.350)       & -1.420(2.550)      & -0.070(0.860)     & -0.310(0.630) & 0.130(0.150)  & 0.270(0.400)     & 0.110(0.240)   \\
         & 0.040(0.350)  & -0.380(0.990)      & 0.300(2.260)       & 0.430(1.410)      & 0.290(0.720)  & 0.110(0.130)  & 0.130(1.210)     & -0.240(0.710)  \\
         & 0.010(0.080)  & -0.090(0.150)      & 0.050(0.150)       & 0.110(0.160)      & 0.120(0.140)  & 0.150(0.100)  & 0.010(0.180)     & -0.090(0.140)  \\
         & 0.010(0.040)  & -0.010(0.050)      & 0.010(0.160)       & -0.010(0.120)     & -0.040(0.120) & 0.020(0.040)  & -0.010(0.110)    & -0.010(0.050)  \\
         & -0.030(0.070) & 0.100(0.150)       & -0.070(0.180)      & -0.100(0.130)     & -0.090(0.150) & -0.170(0.090) & -0.000(0.140)    & 0.100(0.130)   \\ \hline
\end{tabular}
% \end{table}
\end{sidewaystable}

\end{document}